\documentclass[fleqn,usenatbib]{mnras}

\usepackage{newtxtext,newtxmath}

\usepackage[T1]{fontenc}

\DeclareRobustCommand{\VAN}[3]{#2}
\let\VANthebibliography\thebibliography
\def\thebibliography{\DeclareRobustCommand{\VAN}[3]{##3}\VANthebibliography}

\usepackage{graphicx}	
\usepackage{amsmath}	
\usepackage{amssymb}	

\title[SB2 binary TYC~2990-127-1]{TYC~2990-127-1: an Algol-type SB2 binary system of subgiant and red giant  with a probable ongoing mass-transfer.}

\author[M. Kovalev et al.]{
Mikhail Kovalev,$^{1,2,3}$\thanks{E-mail: mikhail.kovalev@ynao.ac.cn}
Zhenwei Li,$^{1,2}$
Xiaobin Zhang,$^{4}$
Jiangdan Li,$^{1,2,5}$
Xuefei Chen$^{1,2,6}$
\newauthor
and Zhanwen Han$^{1,2,6}$
\\
$^{1}$Yunnan Observatories, China Academy of Sciences, Kunming 650216, China\\
$^{2}$Key Laboratory for the Structure and Evolution of Celestial Objects, Chinese Academy of Sciences, Kunming 650011, China\\
$^{3}$Max Planck Institute for Astronomy, D-69117 Heidelberg, Germany\\
$^{4}$Key Laboratory of Optical Astronomy, National Astronomical Observatories, Chinese Academy of Sciences, Beijing, 100012, China\\
$^{5}$University of the Chinese Academy of Sciences, Yuquan Road 19, Shijingshan Block, Beijing 100049, China\\
$^{6}$Center for Astronomical Mega-Science, Chinese Academy of Sciences, 20A Datun Road, Chaoyang District, Beijing 100012, China\\
}

\date{Accepted 22.04.2022. Received 21.04.2022; in original form 17.11.2021}

\pubyear{2022}

\def\kms{\,{\rm km}\,{\rm s}^{-1}}

\def\feh{\hbox{[Fe/H]}}

\newcommand{\teff}{T_{\rm eff}}
\newcommand{\rv}{{\rm RV}}
\def\Vmic{V_{\rm mic}}
\def\Vmac{V_{\rm mac}}
\def\vsini{V \sin{i}}
\def\logg{\log{\rm (g)}}
\def\snr{\hbox{S/N}}
\newcommand{\ha}{\hbox{H$\alpha$}}

\begin{document}
\label{firstpage}
\pagerange{\pageref{firstpage}--\pageref{lastpage}}
\maketitle

\begin{abstract}
We present a study of the spectroscopic binary TYC 2990-127-1 from the LAMOST survey. We use full-spectrum fitting to derive radial velocities and spectral parameters. The high mass ratio indicates that the system underwent mass transfer in the past. We compute the orbital solution and find that it is a very close sub-giant/red giant pair on circular orbit, slightly inclined to the sky-plane. Fitting of the TESS photometrical data confirms this and suggests an inclination of $i\sim 39.8^\circ$. The light curve and spectrum around $\ha$ show signs of irregular variability, which supports ongoing mass transfer. The binary evolution simulations suggest that the binary may experience non-conservative mass transfer with accretion efficiency 0.3, and the binary will enter into common envelope phase in the subsequent evolution. The remnant product after the ejection of common envelope may be a detached double helium white dwarf (He WD) or a merger.
\end{abstract}




\section{Introduction}

Many stars form gravitationally bound systems which can be observed via different astronomical techniques. If a stellar spectrum has composite structure and varies over time it can be produced by system of multiple stars. Such stars are spectroscopic binaries. If clear double lined structure is visible in the spectrum we classify them as an SB2 binary.  
SB2 binaries are very useful astronomical objects that allow us to study many stellar properties. Along with gravitational waves observations \citep{Abbott_2020} and extremely rare microlensing events \citep[][]{singlestarmass,astrometrylens} they allow direct measurement of stellar masses. 
A very special case of the SB2 type are eclipsing binaries (like famous Algol $\beta$Per), where the orientation of the orbit allows us to see periodic eclipses. 
\par
Star TYC 2990-127-1 has been reported in SIMBAD as a "Double or multiple star" with two components in Gaia DR2 \citep{gdr2}: 816182638239124736 ($G=11.3432\pm0.0031$ mag) and 816182638237972736 ($G=13.4266\pm0.0051$ mag) separated by $\sim0.8$ arcseconds. For the first star there is a radial velocity measurement ${\rm RV}=-44.70\pm3.34~\kms$ and  parallax $\varpi=1.4295\pm0.1358$ mas. However, in the recent early Gaia DR3 \citep{egdr3} only one component remains\footnote{within 5 arcsec cone}: 816182638239124736 ($G=11.387562\pm0.002986$ mag) with parallax $\varpi=0.9985\pm0.0343$ mas.
\par
TYC 2990-127-1 is included in the TESS input catalogue \citep[TIC][]{tic} as TIC 56933180\footnote{Since TIC is based on Gaia DR2 this star has two ids: 56933180 and 801622286.} ($T=10.8472$ mag). \citet{kelt} estimated the upper limits for the variability ${\rm rms}\, T= 0.017,\,0.015,\,0.014$ mag for the $30$ min, $2$h and $1$ day exposures, respectively.  
\citet{ting2019} measured spectral parameters and 15 chemical abundances from the APOGEE DR14 \citep{dr14} spectrum of TYC 2990-127-1, although these results are flagged as unreliable with macroturbulence velocity  $\Vmac>20\,\kms$.
This star is absent in the catalog of RV Variable Star Candidates 
from LAMOST (Large Sky Area Multi-Object fiber Spectroscopic Telescope) by \citet{Tian_2020}. 
A recent study by \citet{2021arXiv210710860K} reports TYC 2990-127-1 as a SB2 candidate based on an analysis of cross-correlation functions (CCF) in APOGEE DR16 \citep{dr16}.
\par
In this paper we use LAMOST medium resolution spectra to confirm that TYC 2990-127-1 is a SB2 system of sub-giant and red giant stars, orbiting each other on a circular orbit with period $\sim1$ week.
The paper is organised as follows: in Section~\ref{Observations} we describe the observations and methods. Section~\ref{results} presents our results. In Section~\ref{discus} we discuss the results in context of binary system evolution. In Section~\ref{concl} we summarise the paper and draw conclusions.

\section{Observations \& Methods }
\label{Observations}
\subsection{Observations}

We use the 25 spectra of the TYC 2990-127-1 observed within the LAMOST 
spectroscopic survey \citep{2012RAA....12.1197C, 2015RAA....15.1095L, 2012RAA....12..723Z} with id J091914.65+423348.0. 
We use the spectra taken at a resolving power of R$=\lambda/ \Delta \lambda \sim 7\,500$. Each spectrum is divided on two arms: blue from 4950\,\AA~to 5350\,\AA~and red from 6300\,\AA~to 6800\,\AA. We convert the  heliocentric  wavelength scale in the observed spectra from vacuum to air using \texttt{PyAstronomy} \citep{pya}. The average signal-to-noise ratio ($\snr$) of a spectrum ranges from 23 to 244 ${\rm pix}^{-1}$ for a blue arm and from 65 to 368 ${\rm pix}^{-1}$ for a red arm, with the majority of the spectra sampling the $\snr$ in the range of 100-160 ${\rm pix}^{-1}$. Observations are carried out from 2018-12-14 until 2021-03-24, covering time base of 830 days. During the observations several spectra were taken on consecutive exposures over the night, however we only use the stacked spectrum for whole night due to higher S/N ratio.

\subsection{Spectral models}
The synthetic spectra are generated using the NLTE~MPIA online-interface \url{https://nlte.mpia.de} \citep[see Chapter~4 in][]{disser} on wavelength intervals 4870:5430 \AA~for the blue arm and 6200:6900 \AA ~for the red arm with spectral resolution $R=7500$. We use NLTE (non-local thermodynamic equilibrium) spectral synthesis for H, Mg~I, Si~I, Ca~I, Ti~I, Fe~I and Fe~II lines \citep[see Chapter~4 in][ for references]{disser}.  
\par
The grid of models (6200 in total) is computed for points randomly selected in a range of $\teff$ between 4600 and 8800 K, $\logg$ between 1.0 and 4.8  (cgs units), $\vsini$ from 1 to 300 $\kms$ and [Fe/H]\footnote{We used $\feh$ as a proxy of overall metallicity, abundances for all elements are scaled with Fe.} between $-$0.9 and $+$0.9 dex. The model is computed only if linear interpolation of MAFAGS-OS\citep[][]{Grupp2004a,Grupp2004b} stellar atmosphere is possible for a given point in parameter space.  Microturbulence is fixed to $\Vmic=2~\kms$ for all models. The grid is randomly split on training (70\%) and cross-validation (30\%) sets of spectra, which are used to train \textit{The~Payne} spectral model \citep{ting2019}. The neural network (NN) consists of two layers of 300 neurons each with rectilinear unit (ReLU)\footnote{ReLU(x)=max(x,0)} activation functions. We train separate NNs for each spectral arm. The median approximation error is less than 1\% for both arms. We use output of \textit{The Payne} as a single-star spectral model.

\subsection{Spectral fitting}
\label{sec:maths} 

Our spectroscopic analysis includes two consecutive stages: 
\begin{enumerate}
    \item analysis of individual observations by binary and single-star spectral models, where we normalise the spectra and make a rough estimation of the spectral parameters, see brief description in Section~\ref{sec:ind}. This is a further development of the method presented in \citet{2021arXiv210813853K}. 
    \item simultaneous fitting of multi-epochs with a binary spectral model, using constraints from binary dynamics and values from the previous stage as an input, see Section~\ref{sec:multi}.
\end{enumerate}

\subsubsection{Individual spectra.}
\label{sec:ind}

The normalised binary model spectrum is generated as a sum of the two Doppler-shifted normalised single-star model spectra ${f}_{\lambda,i}$ scaled according to the difference in luminosity, which is a function of the $\teff$ and stellar size. We assume both components to be spherical and use following equation:    

\begin{align}
    {f}_{\lambda,{\rm binary}}=\frac{{f}_{\lambda,2} + k_\lambda {f}_{\lambda,1}}{1+k_\lambda},~
    k_\lambda= \frac{B_\lambda(\teff{_{,1}})~M_1}{B_\lambda(\teff{_{,2}})~M_2} 10^{\logg_2-\logg_1}
	\label{eq:bolzmann}
\end{align}
 where  $k_\lambda$ is the luminosity ratio per wavelength unit, $B_\lambda$ is the black-body radiation  (Plank function), $\teff$ is the effective temperature, $\logg$ is the surface gravity and $M$ is the mass. Throughout the paper we always assume the primary star to be brighter.
\par
The binary  model spectrum is later multiplied by the normalisation function, which is a linear combination of the first four Chebyshev polynomials \citep[similar to][]{kovalev19}, defined separately for blue and red arms of the spectrum. The resulting spectrum is compared with the observed one using \texttt{scipy.optimise.curve\_fit} function, which provides optimal spectral parameters, radial velocities (RV) of each component plus mass ratio $q={M_1}/{M_2}$ and two sets of four coefficients of Chebyshev polynomials. We keep metallicity equal for both components. In total we have 18 free parameters for a binary fit. We estimate goodness of the fit parameter by reduced $\chi^2$:

\begin{flalign}
\label{eq:chi2}
   \chi^2 =\frac{1}{N-18} \sum \left[ \left({f}_{\lambda,{\rm observed}}-{f}_{\lambda,{\rm model}}\right)/{\sigma}_{\lambda}\right]^2
\end{flalign}
where $N$ is a number of wavelength points in the observed spectrum. To explore whole parameter space and to avoid local minima we run the optimisation six times with different initial parameters of the optimiser. We select the solution with minimal  $\chi^2$ as a final result. 
\par 
Additionally, every spectrum is analysed by a single star model, which is identical to a binary model when the parameters of both components are equal, so we fit only for 13 free parameters. Using this single star solution, we compute the difference in reduced $\chi^2$ between two solutions and the improvement factor, computed using Equation~\ref{eqn:f_imp} similar to \cite{bardy2018}. This improvement factor estimates the absolute value difference between two fits and weights it by difference between two solutions.

\begin{align}
\label{eqn:f_imp}
f_{{\rm imp}}=\frac{\sum\left[ \left(\left|{f}_{\lambda,{\rm single}}-{f}_{\lambda}\right|-\left|{f}_{\lambda,{\rm binary}}-{f}_{\lambda}\right|\right)/{\sigma}_{\lambda}\right] }{\sum\left[ \left|{f}_{\lambda,{\rm single}}-{f}_{\lambda,{\rm binary}}\right|/{\sigma}_{\lambda}\right] },
\end{align}
where ${f}_{\lambda}$ and ${\sigma}_{\lambda}$ are the observed flux and corresponding uncertainty, ${f}_{\lambda,{\rm single}}$ and ${f}_{\lambda,{\rm binary}}$ are the best-fit single-star and binary model spectra, and the sum is over all wavelength pixels.

\subsubsection{Multi-epochs fitting.}
\label{sec:multi}

\begin{figure}
    \includegraphics[width=\columnwidth]{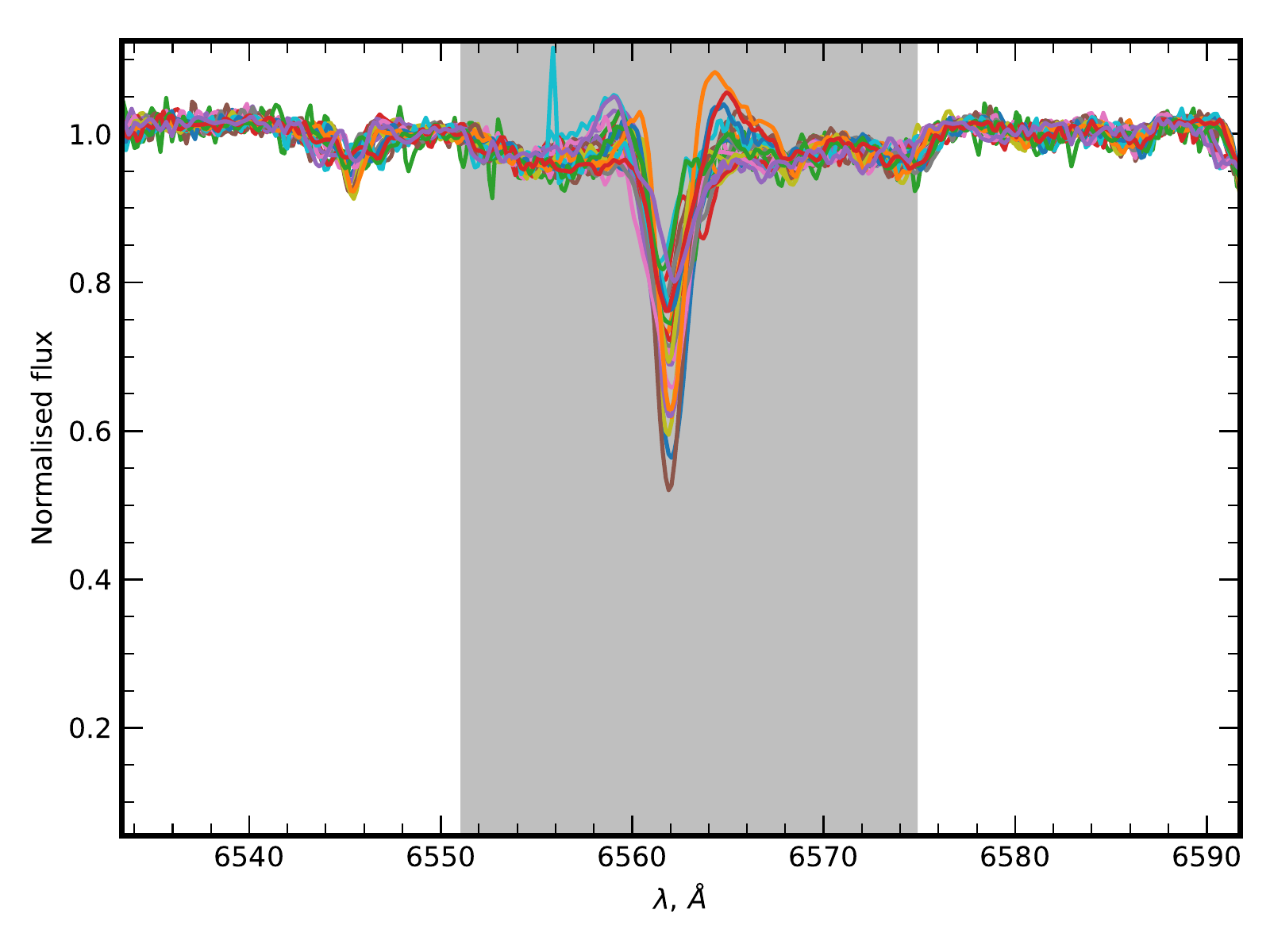}
    \caption{$\ha$ line region in all epochs. The gray shadow shows the masked region excluded from the multi epochs fit.}
    \label{fig:half}
\end{figure}

After visual inspection of the individual epochs fits we find that observed spectra around $\ha$ line were poorly fitted for all epochs. Moreover, there are signs of emission, variable with time, as shown in Figure~\ref{fig:half}. Our models don't allow for the calculation of emission lines,
 therefore we mask out region 12~\AA~around $\ha$ during multi-epochs fitting.  

If two components in our binary system are gravitationally bound, their radial velocities should agree with the following equation \citep{wilson,fernandez2017}:  

\begin{align}
\label{eqn:asgn}
    {\rm RV_{\rm scd}}=\gamma_{\rm dyn} (1+q_{\rm dyn}) - q_{\rm dyn} {\rm RV_{\rm prm}},
\end{align}
where $q_{\rm dyn}={M_{\rm prm}}/{M_{\rm scd}}$ - mass ratio of binary components and $\gamma_{\rm dyn}$ - systemic velocity. Using this equation we can directly measure the systemic velocity and mass ratio. 
\par
\begin{figure}
	\includegraphics[width=\columnwidth]{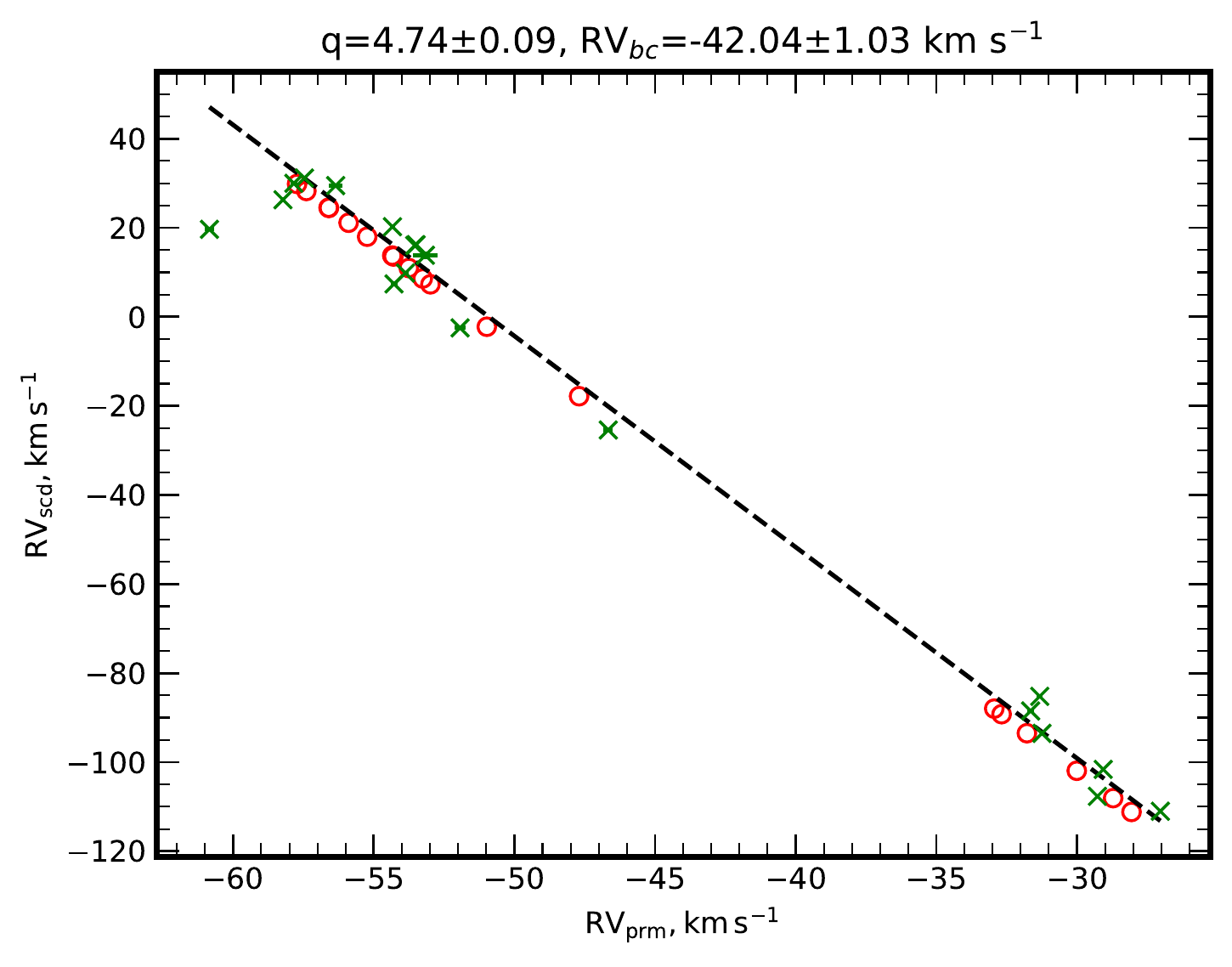}
    \caption{${\rm RV_{\rm prm}}$ vs ${\rm RV_{\rm scd}}$ fit (Wilson plot). The best fit is shown as a black dashed line, RV data from individual epochs used in fitting are shown as green crosses. Corresponding RV data from multiple fit are shown as red circles, they follow a straight line by definition, where only ${\rm RV_{\rm prm}}$ were actually measured and ${\rm RV_{\rm scd}}$ were calculated using Eq.~\ref{eqn:asgn}.}
    \label{fig:rvplot}
\end{figure}

We check radial velocities derived from individual spectra and find that for several epochs the secondary radial velocity ${\rm RV_{\rm scd}}$ is very different from the primary radial velocity ${\rm RV_{\rm prm}}$ (see Table~\ref{tab:rvs}), while ${\rm RV_{\rm prm}}$ is very close to the radial velocity estimate from the single star model. Visual inspection of such spectra has confirmed that they have the wrong ${\rm RV_{\rm scd}}$ as the actual radial velocity separation is close to zero, the regime where binary model fails. We select all remaining 19 RV data points and fit a line via Equation~\ref{eqn:asgn}, using orthogonal distance regression (ODR) method \citep{odr}\footnote{ \url{https://docs.scipy.org/doc/scipy/reference/odr.html}} which can handle uncertainty in both variables.  
In Figure~\ref{fig:rvplot} (Wilson plot) we present the resulting best fit as a dashed line and data points as green crosses. Note the absence of data points near the systemic velocity -- these six data points are excluded.
\par
To reduce the number of free parameters in multi-epochs fitting we use only ${\rm RV_{\rm prm}}$ and computed ${\rm RV_{\rm scd}}$ using Equation~\ref{eqn:asgn}. The same value of mass ratio is used in Equation~\ref{eq:bolzmann}. Unlike the previous stage, we fit $\feh$ for both components. In total, we fit for 35 free parameters.   
We use the previously derived $\gamma_{\rm dyn}$, $q_{\rm dyn}$ and ${\rm RV_{\rm prm}}$ as initial guesses for these parameters. We fit all previously normalised individual epoch's spectra three times using spectral parameters with three maximal improvement factor values for initialisation. We select the solution with minimal  $\chi^2$ as a final result.  

\subsubsection{Typical errors estimation}
\label{sec:err}
We estimate typical errors of the multi-epochs fitting by testing it's performance on the dataset of synthetic binaries. We generated 3000 mock binaries using uniformly distributed mass-ratios from 2.5 to 10.0, $\teff$ from $4600$ to $6300$ K, $\logg$ from 1.0 to 3.0 (cgs) and $\vsini$ from 1 to 100 $\kms$. Metallicity was set to $\feh=-0.6$ dex for both components. For each star, we computed 5 mock binary spectra using radial velocities computed for circular orbits with the semiamplitude of the primary component $20\,\kms$ at randomly chosen phases. These models were degraded by Gaussian noise according to $\snr=100$ pix$^{-1}$. 
We performed exactly the same analysis as for the observations on this simulated dataset. We checked how well the mass ratio and the spectral parameters of the primary and secondary components can be recovered by calculating the average and standard deviation of the residuals. For the primary components we have $\Delta\teff=2\pm122$~K, $\Delta\logg=0.01\pm0.08$ cgs units, $\Delta\vsini=0\pm7\kms$  and $\Delta\feh=0.00\pm0.04$ dex. For the secondary components we have $\Delta\teff=-21\pm319$~K, $\Delta\logg=0.00\pm0.18$ cgs units, $\Delta\vsini=-6\pm29\kms$ and $\Delta\feh=0.04\pm0.18$ dex. The mass ratio recovery has $\Delta q=-0.02\pm0.09$. It is clear that the parameters of the secondary components are poorly recovered compared to the primary components and there are no significant biases in the recovered results.

\subsection{Orbital fitting}

In the next step, radial velocities ${\rm RV_{\rm prm}}$ are used to fit circular orbits using the generalised Lomb-Scargle periodogram (\texttt{GLS}) code by \cite{gls} :

\begin{align}
    {\rm RV}(t)=\gamma- K \sin \left (\frac{2\pi}{P}(t -t_0) \right ),
\end{align}%
where $\gamma$ - systemic velocity, $P$ - period, $t_0$ -conjunction time, $K$ - radial velocity semiamplitude.
We also fit to a Keplerian orbit and find that eccentricity is equal to zero, so a circular orbit is a valid assumption. Now we can use the period, binary mass ratio and the radial velocity semiamplitude to compute the radius of the orbit $a=(a_{\rm prm}+a_{\rm scd})$ and to constrain the total mass $M_{\rm tot}=(M_2+M_1)$ using Kepler's third law:

\begin{align}
    a\sin i=(K +q K)\frac{P}{2 \pi},~M_{\rm tot}\sin^3{i}=\frac{(K+q K)^3}{GM_{\odot}} \frac {P}{2 \pi} ,
    \label{eq:kepler3}
\end{align}
where $GM_\odot=1.32712440041\cdot 10^{20}\, {\rm m^3\,s^{-2}}$  is the Solar mass parameter\footnote{\url{https://iau-a3.gitlab.io/NSFA/NSFA_cbe.html\#GMS2012}}, $i$ is the inclination of the orbit with respect to the sky-plane. Masses of the components can be found using total mass and $q$. 

\begin{figure*}
    \includegraphics[width=\textwidth]{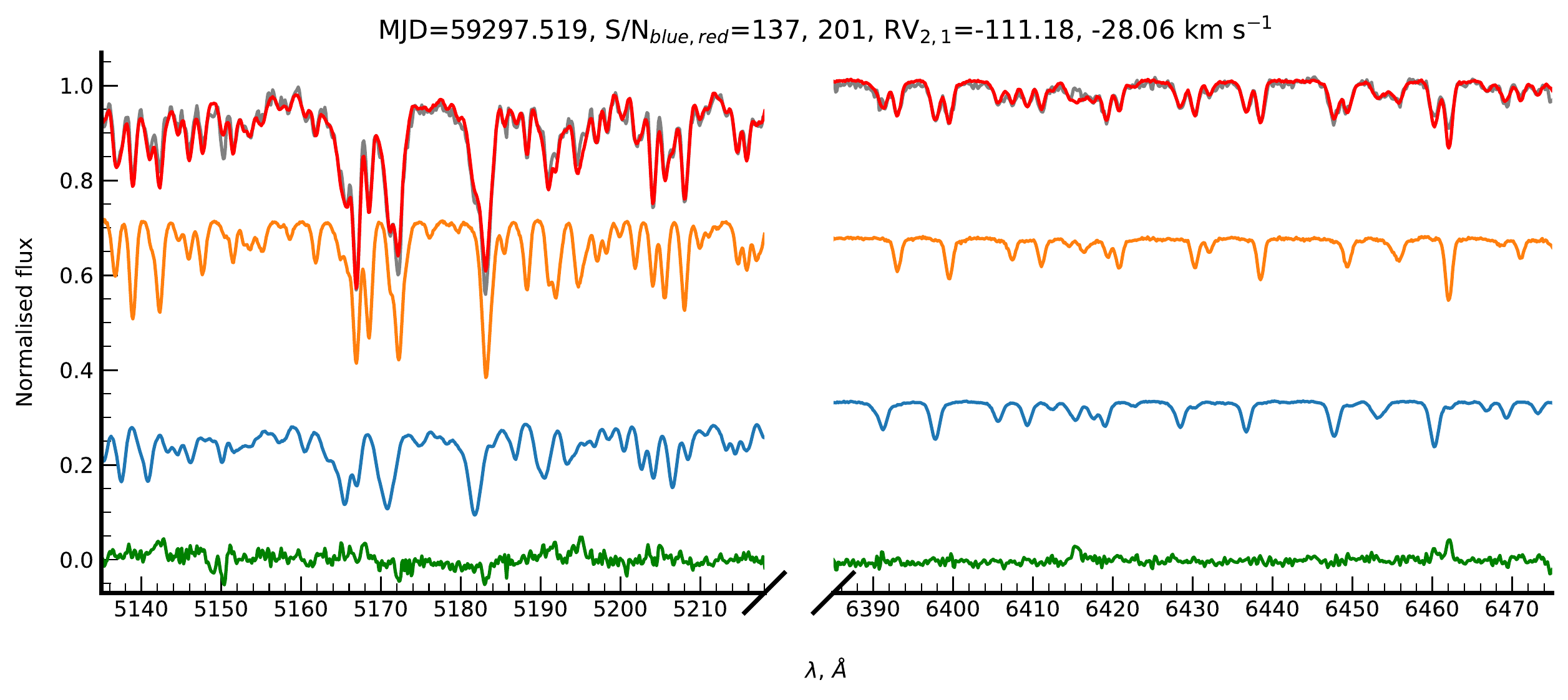}
    \caption{Example of the spectrum fitting for epoch with maximal radial velocity separation. The observed spectrum is shown as a gray line, the best fit is shown as red line. The primary component is shown as the orange line, the secondary as a blue line. The difference O-C is shown as a green line.}
    \label{fig:spfit}
\end{figure*}

\section{Results}
\label{results}
In Figure~\ref{fig:spfit} we show the best fit by multi-epochs binary model for the epoch with maximal RV separation. We zoom into the wavelength range around the magnesium triplet in the blue arm and in a 90~\AA~interval in the red arm, where many double lines are clearly visible. The primary star is the subgiant ($\teff=5627~K, \logg=3.48$ cgs, $\feh=-0.67$ dex, $\vsini=21~\kms$) which contributes around 70\% of the visible light, while the secondary star is the red giant ($\teff=4727~K, \logg=2.78$ cgs, $\feh=-0.46$ dex, $\vsini=31~\kms$) which contributes to the remaining 30\%. The errors in the spectral parameters provided by \texttt{scipy.optimise.curve\_fit} are nominal and largely underestimated, so we omit them. 
\par 
The derived mass ratio $q=4.75$ and systemic radial velocity $\gamma=-42.51~\kms$ agree well with values derived by ODR: $q_{\rm dyn}=4.74\pm0.09$, $\gamma_{\rm dyn}=-42.04\pm1.03~\kms$. Surface gravities and mass ratio allow us to estimate ratio of component's sizes $R_{\rm prm}=0.97R_{\rm scd}$. Figure~\ref{fig:rvplot} shows the radial velocities from fitting multiple spectra; they follow a straight line by definition as ${\rm RV_{\rm scd}}$ are computed with Equation~\ref{eqn:asgn}.  

\par

\begin{figure}
	\includegraphics[width=\columnwidth]{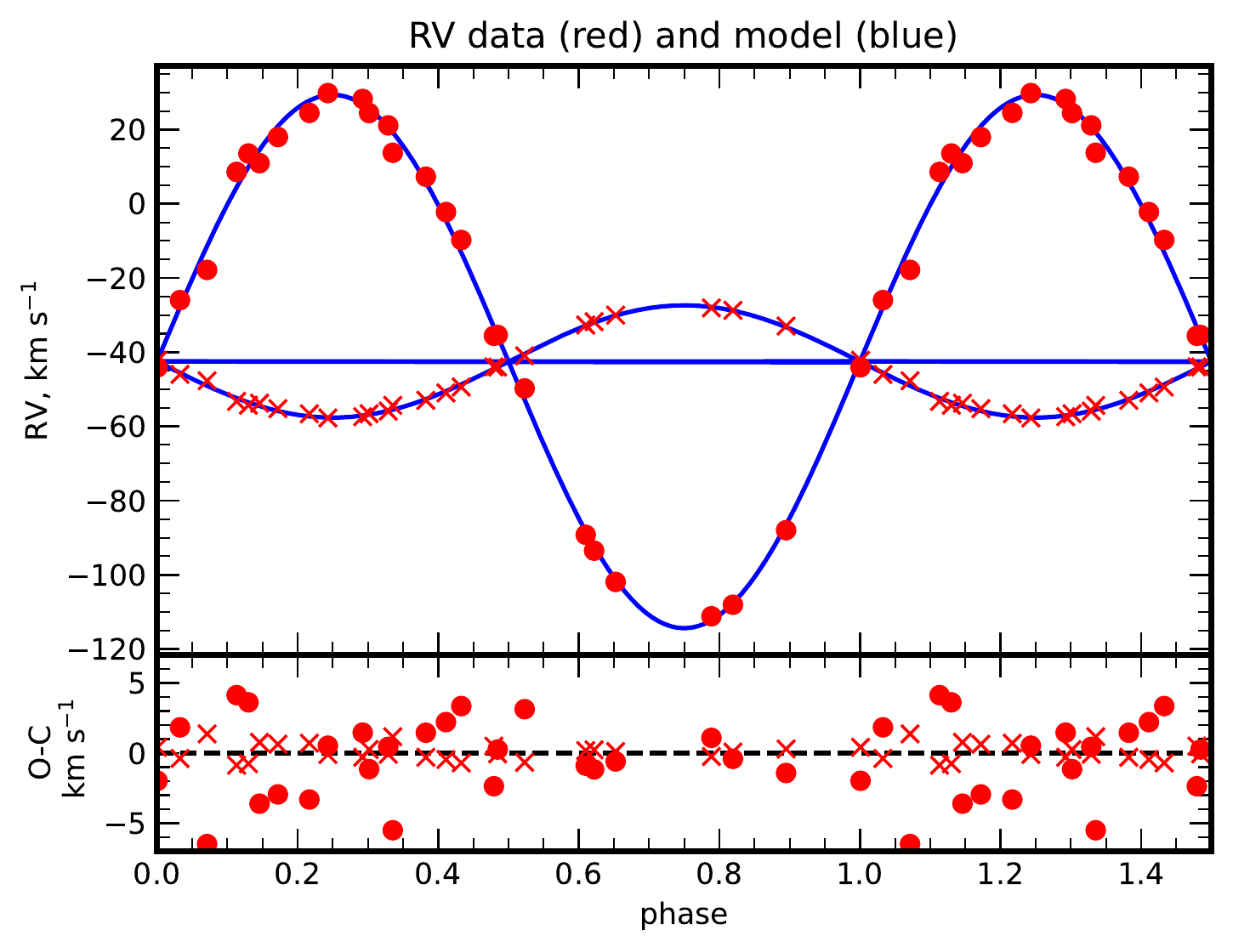}
    \caption{Circular orbit fit with \texttt{GLS} (phase folded). Primary RV curve (measured) is shown by crosses, secondary RV curve (calculated using RV primary, $\gamma$ and the mass ratio) is shown as circles. In the bottom panel we show fit residuals O-C.}
    \label{fig:orbit_rg}
\end{figure}

In Figure~\ref{fig:orbit_rg}, we show the best fit of a circular orbit by \texttt{GLS} (blue line) to the ${\rm RV_{\rm prm}}$ from multi-epochs analysis (red crosses). The fit residuals are small ($\leq 1~\kms$). Note that we also show the secondary star orbit and ${\rm RV_{\rm scd}}$ (red circles), however these data were not used for fitting. We show them to better illustrate orbital motion in binary system. The orbit is well-characterised, such that conjunctions and RV extreme phases are covered. The derived systemic velocity $\gamma=-42.51 \pm 0.13\, \kms$ is equal to one derived from multi-epoch spectral fit.  

\begin{table}
    \centering
    \caption{Radial velocity measurements. The asterix (*) denotes datapoints which are not used in ODR fit}
    \begin{tabular}{cccc}
\hline
\hline
 & multi-epochs&  ind. epoch & single  \\
HJD & RV$_{\rm prm}$ & RV$_{\rm prm}$ , RV$_{\rm scd}$&RV \\
-2400000 d   & $\kms$ & $\kms$ & $\kms$\\

\hline
 58467.35 &-47.70$\pm$0.03 & -46.66$\pm$0.17,  -25.40$\pm$0.26 & -37.10  \\
 58496.28 &-55.23$\pm$0.03 & -54.33$\pm$0.08,   20.23$\pm$0.16 & -32.51  \\
 58510.21 &-53.75$\pm$0.11 & -53.16$\pm$0.43,   13.83$\pm$0.82 & -30.45  \\
 58511.25 &-57.39$\pm$0.04 & -57.84$\pm$0.12,   29.98$\pm$0.20 & -20.99  \\
 58512.24 &-49.40$\pm$0.03 & -38.10$\pm$0.09,  223.50$\pm$0.76*& -36.70  \\
 58560.11 &-56.61$\pm$0.07 & -56.35$\pm$0.24,   29.47$\pm$0.45 & -32.19  \\
 58801.40 &-50.98$\pm$0.04 & -51.93$\pm$0.20,   -2.48$\pm$0.39 & -33.80  \\
 58821.39 &-57.73$\pm$0.02 & -57.46$\pm$0.04,   31.17$\pm$0.06 & -21.12  \\
 58823.36 &-40.98$\pm$0.03 & -44.17$\pm$0.04,   18.16$\pm$0.55*& -44.09  \\
 58831.33 &-30.01$\pm$0.03 & -29.07$\pm$0.06, -101.63$\pm$0.09 & -54.32  \\
 58836.15 &-54.34$\pm$0.02 & -53.49$\pm$0.05,   16.21$\pm$0.08 & -29.32  \\
 58851.32 &-44.01$\pm$0.03 & -40.93$\pm$0.02,  -66.20$\pm$0.36*& -40.82  \\
 58852.28 &-31.78$\pm$0.03 & -31.24$\pm$0.07,  -93.54$\pm$0.12 & -54.14  \\
 58859.26 &-32.68$\pm$0.03 & -31.65$\pm$0.13,  -88.49$\pm$0.23 & -53.54  \\
 58861.27 &-32.94$\pm$0.03 & -31.32$\pm$0.10,  -85.23$\pm$0.18 & -51.23  \\
 58883.18 &-42.18$\pm$0.03 & -43.01$\pm$0.03, -114.00$\pm$0.96*& -43.02  \\
 58910.13 &-28.71$\pm$0.03 & -29.27$\pm$0.06, -107.68$\pm$0.09 & -56.82  \\
 58914.10 &-52.98$\pm$0.03 & -54.28$\pm$0.12,    7.40$\pm$0.20 & -33.01  \\
 58943.01 &-43.96$\pm$0.03 & -41.28$\pm$0.03,  280.00$\pm$0.36*& -41.09  \\
 58949.01 &-55.89$\pm$0.05 & -60.84$\pm$0.17,   19.65$\pm$0.26 & -29.45  \\
 59187.41 &-53.26$\pm$0.03 & -53.87$\pm$0.12,    9.85$\pm$0.20 & -33.13  \\
 59236.23 &-45.99$\pm$0.03 & -39.05$\pm$0.04, -252.46$\pm$0.67*& -39.35  \\
 59245.19 &-56.59$\pm$0.04 & -58.23$\pm$0.10,   26.30$\pm$0.17 & -27.45  \\
 59265.15 &-54.30$\pm$0.02 & -53.54$\pm$0.06,   16.02$\pm$0.10 & -32.61  \\
 59298.02 &-28.06$\pm$0.03 & -27.03$\pm$0.07, -111.03$\pm$0.11 & -58.35  \\
\hline
\hline
    \end{tabular}
    \label{tab:rvs}
\end{table}

\par
\subsection{Tidal synchronisation}
The small period, circular orbit and relatively fast rotation of the components allow us to assume tidal synchronisation in the system. In this case, components have the same spin periods and parallel rotation axes, their sizes are related as rotational velocities. Thus we can derive $\logg$ of one component using following relations:

\begin{align}
    \frac{2\pi}{P}R_1\sin{i}=(\vsini)_1,\,\frac{2\pi}{P}R_2\sin{i}=(\vsini)_2,\\
    \logg_1=\logg_2-\log{\frac{M_2}{M_1}}-2\log{\frac{(\vsini)_2}{(\vsini)_1}}
    \label{eq:tidal}
\end{align}
where inclinations $i$ are presumed to be equal to the inclination of the orbit. With this relation we repeat the multi-epochs fitting using a fixed value of the mass ratio. We find that the new spectral parameters are very close to the unconstrained solution, except for $\vsini$, which becomes almost identical for both components. In this case, the ratio of component's sizes $R_{\rm prm}=0.96R_{\rm scd}$, is very similar to the unconstrained solution. We present a compilation of the orbital and spectral parameters in Table~\ref{tab:final}.

\begin{table}
    \centering
    \caption{Orbital and spectral parameters.}
    \begin{tabular}{ccc}
\hline
    & free $\logg_{\rm scd}$ & tidal synchronisation\\
\hline
$P$, days & \multicolumn{2}{c}{$7.0565 \pm 0.0004$}\\
$t_0$, HJD & \multicolumn{2}{c}{$2458473.9284 \pm 0.0138$}\\
$K_{\rm prm},\,\kms$ & \multicolumn{2}{c}{$15.11 \pm 0.19$}\\
$\gamma,\, \kms$ & \multicolumn{2}{c}{$-42.51 \pm 0.13$}\\
$q_{\rm dyn}$ & \multicolumn{2}{c}{$4.74 \pm 0.09$}\\
$M_{\rm tot} \sin^3{i},\, M_\odot$ & \multicolumn{2}{c}{$ 0.48 \pm 0.03$}\\
$a \sin{i},\,R_\odot$ & \multicolumn{2}{c}{$ 12.09 \pm 1.08$}\\
$\teff{}_{\rm prm}$, K & 5627 & 5592 \\
$\logg_{\rm prm}$, cgs & 3.48 & 3.45\\
$\feh_{\rm prm}$, dex & -0.67  & -0.68\\
$\vsini_{\rm prm},\,\kms$ & 21  & 24\\
$\teff{}_{\rm scd}$, K & 4727 & 4769 \\
$\logg_{\rm scd}$, cgs & 2.78 & 2.81\\
$\feh_{\rm scd}$, dex & -0.49 & -0.41\\ 
$\vsini_{\rm scd},\,  \kms$ & 31 & 25\\
\hline
    \end{tabular}
    \label{tab:final}
\end{table}

\subsection{Verification using APOGEE DR16 spectrum.}

An independent observation can be a great test for our spectral parameters and orbital solution.
We find one infrared high resolution spectrum (R$\sim22\,500$) of TYC 2990-127-1 in the APOGEE~DR16 \citep[][]{dr16}. It has APOGEE id=2M09191465+4233479, high S/N=246~${\rm pix^{-1}}$ and was observed on 2014-11-28 (MJD=56989.446 days) more than 1477 days before first observation in our LAMOST set, which make it a perfect test for our orbital solution. \cite{2021arXiv210710860K} reports ${\rm RV_{\rm scd}}=-102.33\pm1.28 \kms$,  ${\rm RV_{\rm prm}}=-31.88\pm1.72 \kms$, which is very close to our orbital solution ${\rm RV_{\rm scd}}=-111.46 \kms$,  ${\rm RV_{\rm prm}}=-28.00 \kms$. The discrepancy is probably due to the different $\rv$ zero points of the spectrographs used in LAMOST and APOGEE \citep[][]{songK2}. 
\par
\begin{figure}
	\includegraphics[width=\columnwidth]{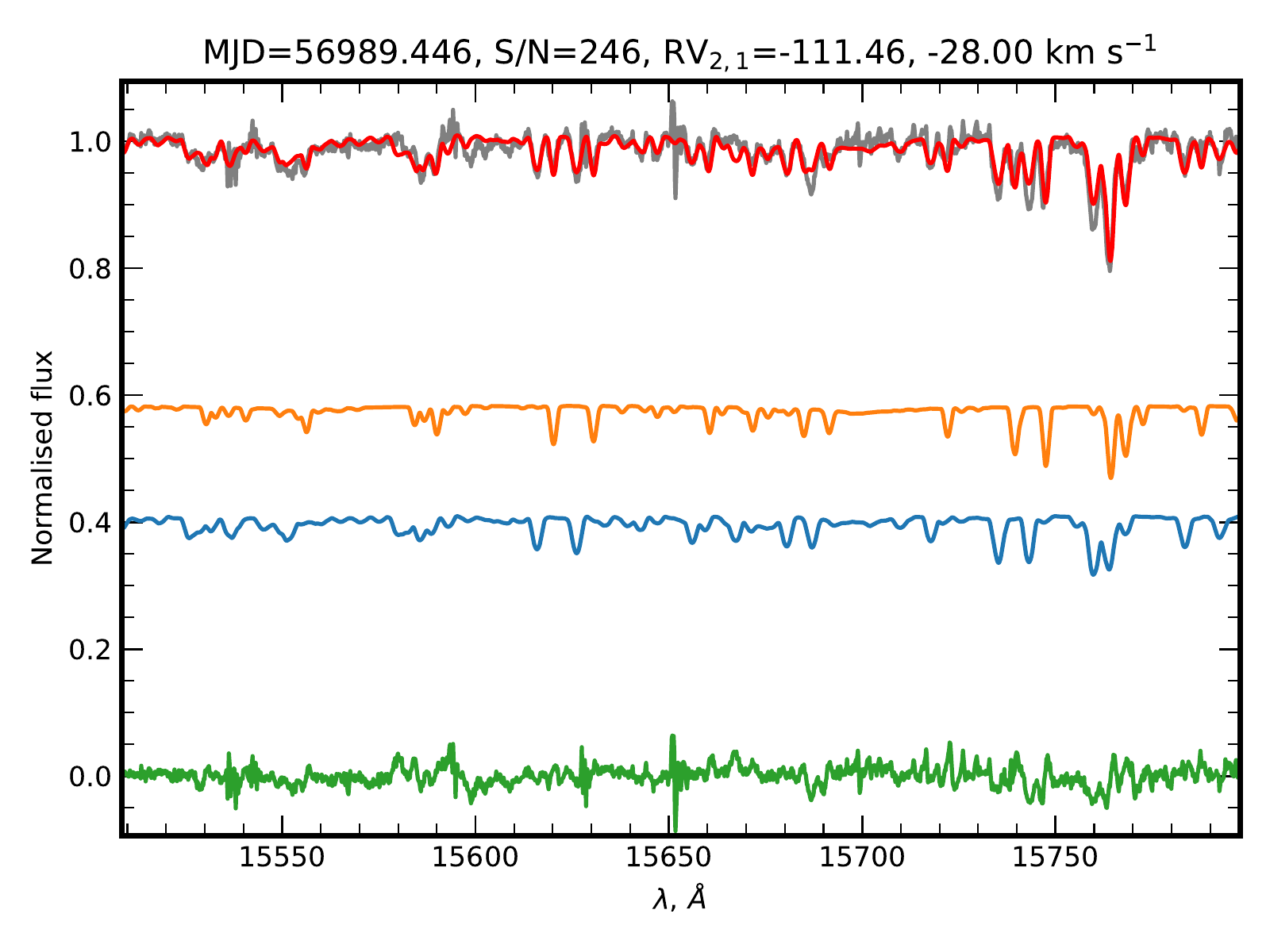}

    \caption{Part of the normalised APOGEE~DR16 spectrum together with the binary model. The primary and secondary component's contributions are also shown. All colours are the same as in Figure~\ref{fig:spfit}.}
    \label{fig:apogee}
\end{figure}

We convert the observed spectrum's wavelength scale from vacuum to air and apply a heliocentric correction using \texttt{PyAstronomy} \citep{pya}.   
We compute two infrared model spectra of the primary and secondary components with the NLTE MPIA\citep{NLTE_MPIA} online interface using our estimations of the spectral parameters and R$=22\,500$. We apply a Doppler shift according to orbital solution for the MJD=56989.446 date and plot the scaled sum of the synthetic spectra in Figure~\ref{fig:apogee} together with the normalised observation. Note that we keep all spectral parameters and radial velocities fixed. We can see that our model reproduces the observation very well, therefore we are confident that our orbital solution is correct. Additionally, the orbital period didn't  change significantly during the last 7 years. We include detailed analysis of this high-resolution spectrum in Section~\ref{sec:payne}. 

\subsection{Light curve from TESS}
\label{tess}

\begin{figure}

	\includegraphics[width=\columnwidth]{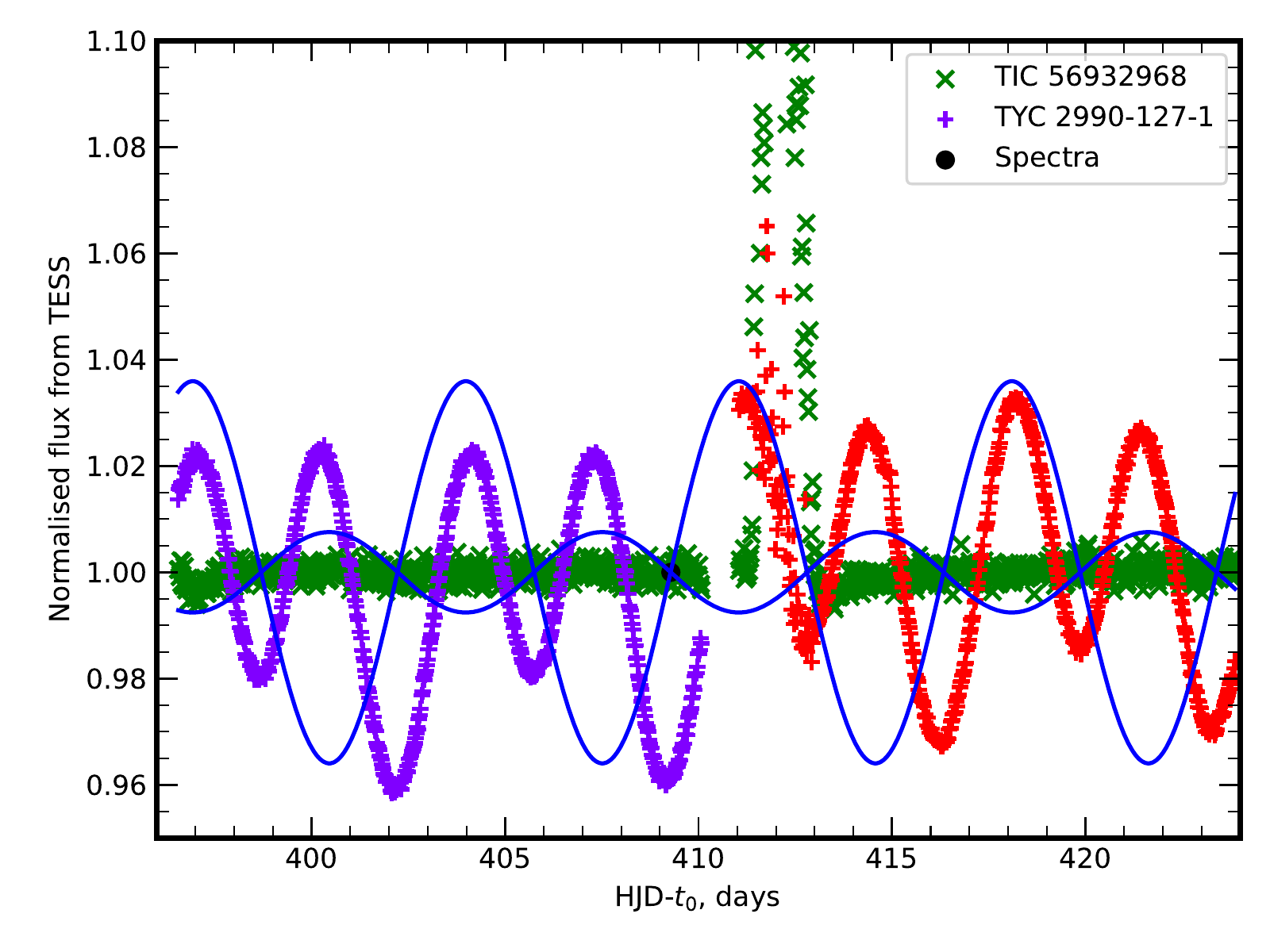}

    \caption{ Normalised TESS light curves for TYC 2990-127-1 and nearby star together with schematic plot of the orbital solution (the solid lines). The black circle shows the moment when one LAMOST spectrum was observed. Datapoints from TESS orbits with numbers 49 and 50 are shown as magenta and red pluses respectively.}
    \label{fig:tess}
\end{figure}

The light curve (LC) from the first release of the TESS photometry \citep{tess1,tess2} covers a timebase of $27.3$ days (almost four complete binary orbits), observed during two TESS orbits with numbers $49$ and $50$ and shows smooth, periodic light variations, see Figure~\ref{fig:tess}.

The period is in clear agreement with our orbital solution. The LC from TESS orbit number 50 shows some problems with contamination at the beginning, which is also visible in the LC of nearby star TIC 56932968, analysed by the same pipeline. Also it is clearly visible that LC of TYC 2990-127-1 shows irregular variability, which is not present in the LC of other star, thus this cannot be explained by data processing issues. Therefore we decide to analyse only LC from orbit number 49, as it is relatively stable (at $\pm0.3\%$ level) over two periods, although the LC is slightly brighter after the second main minimum. 

\par

\begin{figure}
    \includegraphics[width=\columnwidth]{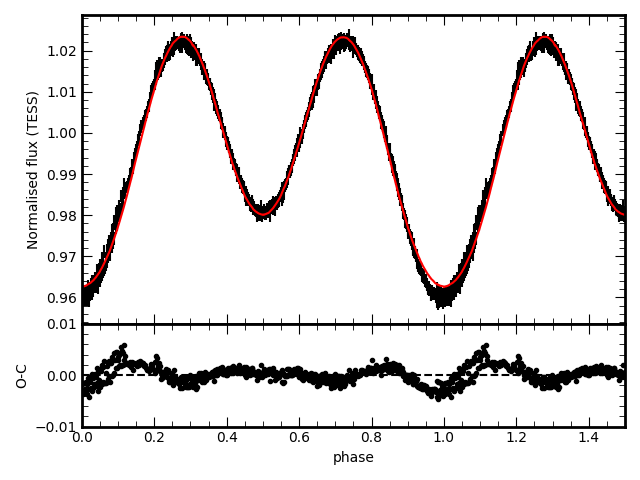}
    \caption{ LC fitting using the \texttt{PHOEBE model}. In the first panel, the black errorbars are the observational phased light curve from TESS, the red line is the best fitting model using two-star model. The second panel shows the residual between the observation and the two-star model.}
    \label{fig:lcnospot}
\end{figure}

\begin{figure}
    \includegraphics[width=\columnwidth]{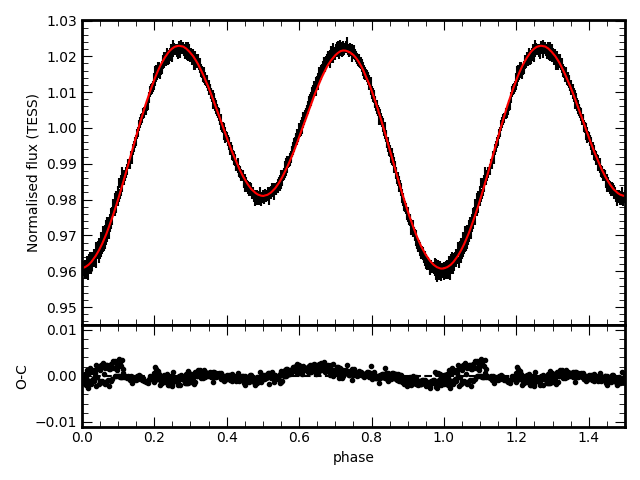}
    \caption{ Same as Figure~\protect\ref{fig:lcnospot}, but for model with spot on the primary.}
    \label{fig:lcspot}
\end{figure}

To model the LC we use the PHysics Of Eclipsing BinariEs (\texttt{PHOEBE}) \citep[][]{phoebe} in a semi-detached binary configuration, where variation of the LC is primarily caused by the distortion of the secondary due to the gravitational force.
We fix the mass ratio $Q=1/q=M_{2}/M_{1}=0.21$, assign $\teff$ based on spectral solution with tidal synchronisation and add a constraint $R_{1}=0.96R_{2}$. We use default \texttt{'ck2004'} atmospheres to let \texttt{PHOEBE} interpolate limb darkening coefficients on the fly. 
 We use the default optimiser to find the best fitting model, minimizing the residual between the model and the observation.
The fitting parameters include an orbital inclination ($i$), $t_0$, albedos ($A_{1,2}$), gravitational darkening exponents ($\alpha_{1,2}$) and $a \sin{i}$. The best fitting solution is provided in  Table~\ref{tab:lc_phoebe}. 
Generally fit is good, although $A,\alpha$ have unusual values and the LC model fails around the phase 0 (main minimum),
as it is shown in Figure~\ref{fig:lcnospot}.
The light curve shows a deep minimum that two stars alone cannot match.
The residuals suggest that light is partially blocked by something i.e. mass flow and/or circumbinary material.
 Since there is no option for including such structures in the \texttt{PHOEBE},
we add a low-temperature spot (with four parameters: latitude, longitude, radius and temperature) on the primary to mimic the decreasing flux.
We use the same initial parameters as for model without a spot, but now we fix $A$ and $\alpha$ to the standard values. The solution with spot is listed in Table~\ref{tab:lc_phoebe}. 
We find that the spot is relatively small (its radius 25 deg), it has an effective temperature $\sim 8\%$ cooler and it is located almost directly opposite to the secondary\footnote{We also found out that similar spot on the secondary, facing $L_2$ point will provide a good fit as well}.
 Given the limitations of the model and the necessary addition of a spot, we take the solution from the optimiser (without exploring the parameter uncertainties, for example, with a sampler) and simply acknowledge that the derived parameters are highly uncertain.
The modelled LC with such a spot and the residuals are presented in  Figure~\ref{fig:lcspot},
where we can clearly see that, with the cold spot, the modelled LC  matches the observations better, although there is small missmatch around the main minimum, probably due to change of the mass flow intensity. Nevertheless, the residuals of the fit are comparable to the level of stability of the LC from TESS orbit 49.

\begin{table}
\centering
\caption{LC solution of TYC 2990-127-1}
\begin{tabular}{lcc}

\hline
\hline
Parameter & Star1 & Star2\\
\hline
$T$, K            & 5592* & 4769*\\
$Q$            &\multicolumn{2}{c}{0.21*}\\
$e$ & \multicolumn{2}{c}{0.0*}\\
$P$, days & \multicolumn{2}{c}{$7.0565*$}\\
\hline
Solution without spot:\\
$i$ , deg     &\multicolumn{2}{c}{49.2}\\
$a \sin{i},\,R_\odot$ & \multicolumn{2}{c}{$ 11.38$}\\
$t_0$, HJD & \multicolumn{2}{c}{$2458473.9522$}\\
$A$            & 0.14  & 0.98\\
$\alpha$            & 0.41 & 0.05\\
Physical parameters (computed):\\
$M,\,M_{\odot}$ & 0.76  & 0.16\\
$R_{\rm equiv},\, R_{\odot}$ & 3.67 & 3.82\\
$\logg$, (cgs)   & 3.19 & 2.48\\
$\vsini,\,\kms$ & 19.93 & 20.76\\
\hline
Solution with spot:\\
$i$  deg     &\multicolumn{2}{c}{39.8}\\
$a \sin{i},\,R_\odot$ & \multicolumn{2}{c}{$ 12.20$}\\
$t_0$, HJD & \multicolumn{2}{c}{$2458473.9775$}\\
$A$            & 0.5*  & 0.5*\\
$\alpha$            & 0.32* & 0.32*\\
Spot parameters:\\
Latitude, deg   & 93.5& {}\\
Longitude, deg   &   7.5& {}\\
Radius, deg      & 25.6& {}\\
$T_{\rm spot}/T_{1}$  & 0.921& {}\\
Physical parameters (computed):\\
$M,\,M_{\odot}$ & 1.54  & 0.32\\
$R_{\rm equiv},\, R_{\odot}$ & 4.65 & 4.85\\
$\logg$, (cgs)   & 3.29 & 2.58\\
$\vsini,\,\kms$ & 21.36 & 22.45\\
\hline
*: assigned.
\end{tabular}
\label{tab:lc_phoebe}
\end{table}

\section{Discussion}
\label{discus}
\begin{figure}
	\includegraphics[width=\columnwidth]{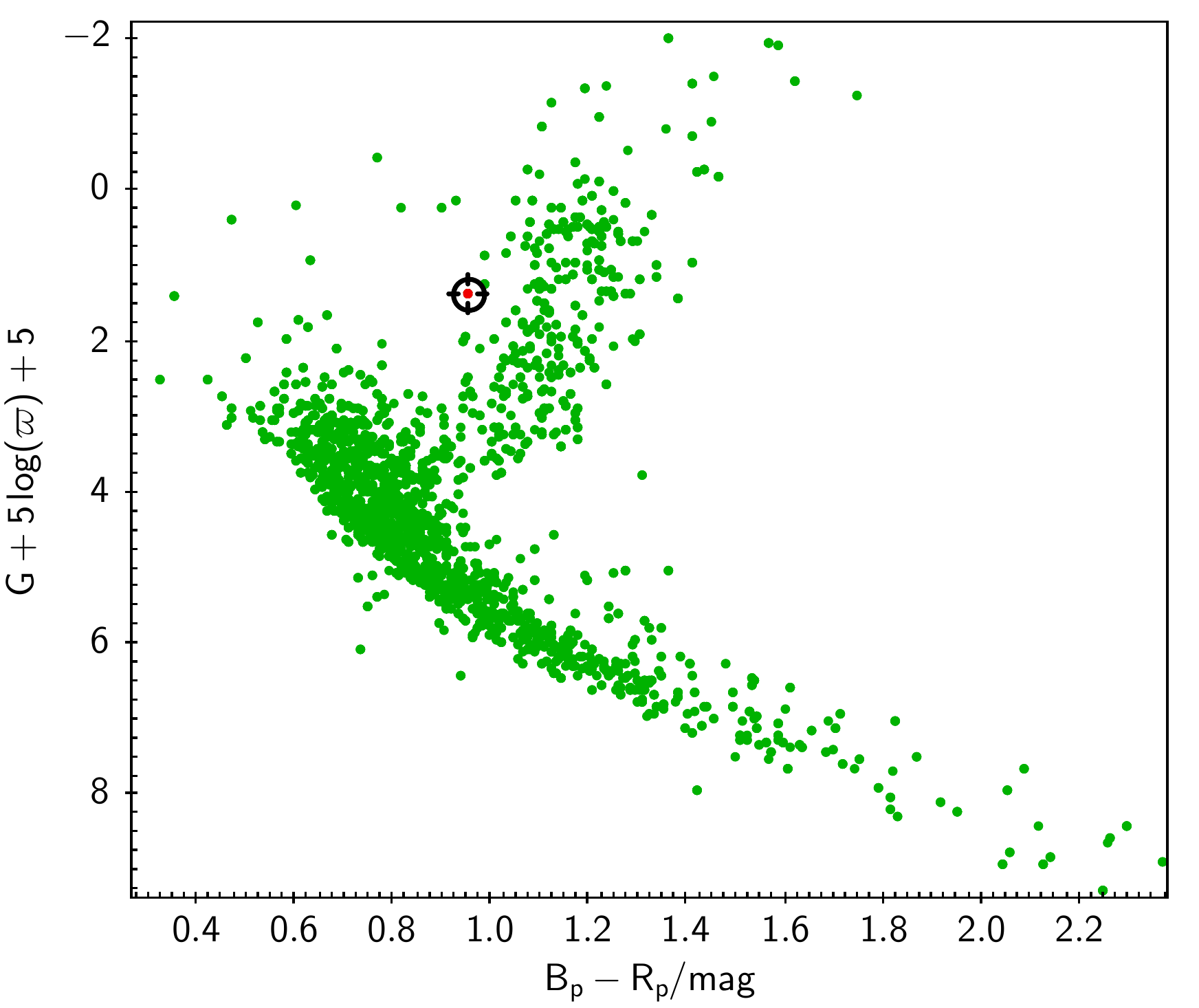}
    \caption{Hertzsprung-Russell diagram for all LAMOST MRS stars in same field ($\sim4^\circ$) with TYC 2990-127-1 (red circle inside target) based on \protect\citet{egdr3} data.}
    \label{fig:hrd_egdr3}
\end{figure}

\begin{figure}
	\includegraphics[width=\columnwidth]{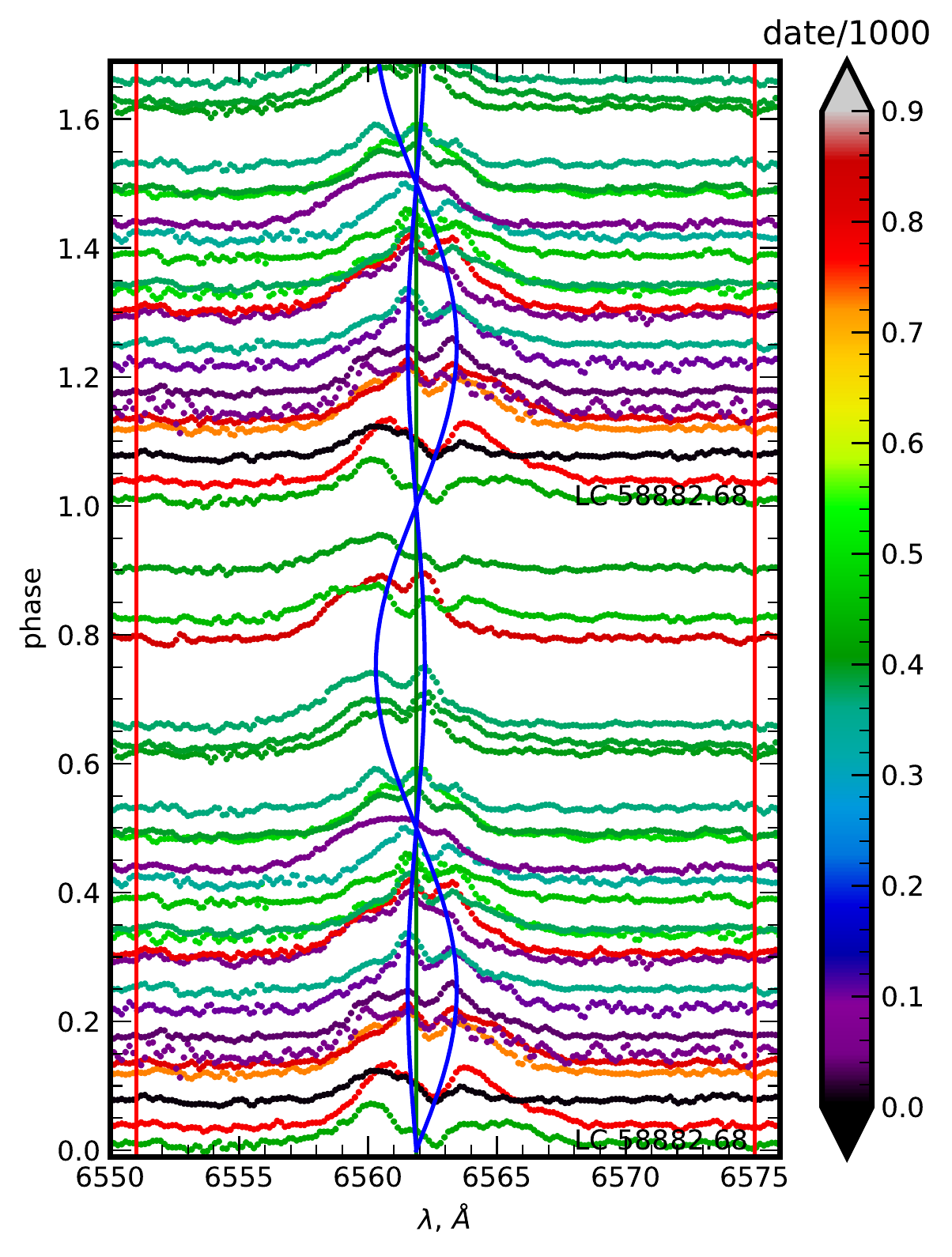}
    \caption{Dynamic spectrum of the $\ha$ line based on the residual profiles after the binary model (solution with free $\logg_{\rm scd}$) has been subtracted. The blue lines are representing positions of $\ha$ line center according to the orbital solution. The red lines are the edges of the mask. The residual value is shown with offset according to orbital phase. The colour of the profiles indicates date of the observation (MJD-58460)/1000. Profile taken during TESS observations marked by label ``LC 58882.68".}
    \label{fig:rezid}
\end{figure}

\begin{figure}
	\includegraphics[width=\columnwidth]{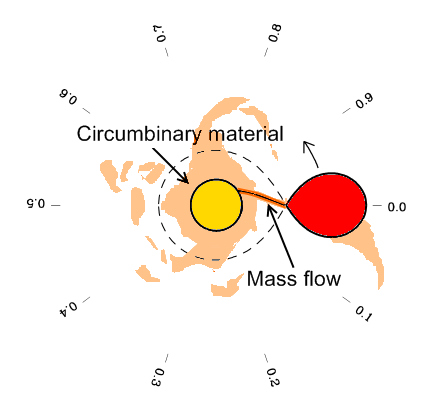}
    \caption{ Schematic picture with the possible explanation of dynamic spectrum. Based on Figure 1 from \protect\citet{richards1999} and Figure 17 from \protect\citet{lehmann2020}. Viewing angles corresponding for different phases are shown around the system. Plot is centered in primary component, with secondary orbiting counterclockwise. }
    \label{fig:cartoon}
\end{figure}

\subsection{System parameters}

In Figure~\ref{fig:hrd_egdr3} we show the Hertzsprung-Russell diagram for all LAMOST MRS stars in same field with TYC 2990-127-1 based on \citet{egdr3} data. We see that the binary is located higher than typical subgiants and to the left of the red giants, which qualitatively confirms our estimation of the spectral parameters. 
\par

 We show the dynamic spectrum of the $\ha$ line in Figure~\ref{fig:rezid}. Here we subtracted the binary model (solution with free $\logg_{\rm scd}$) from the normalized observations and ordered them according to orbital phase.
Emission lines in the $\ha$ region are relatively broad, most clearly associated with both stellar components, but some have significant $\rv$ shifts relative to both components e.g. phases 0.03, 0.42. The profile taken during TESS observations near the main minimum of the LC, shows broad blue-shifted and weaker, much broader red-shifted emission lines. Some strong emission features are probably short-lived, as they disappear with time in the spectra taken at similar phases e.g. phase 0.33.  
 These time-variable emission features are probably due to the non-stationary mass transfer and mass loss events in the system. In Figure~\ref{fig:cartoon} we give a possible explanation of a dynamic spectrum, based on Figure 1 from \citet{richards1999}. The mass flow from the secondary can be be observed near the phases 0.6-0.7 as a broad blue-shifted emission, similarly to Figures 23,24 from \citet{richards1999}. If not accreted completely, this material can form a circumbinary medium, which can partially block the light from the primary and secondary. This can explain why the LC shows a deficiency of light from the primary and best-fit requires cool spot to model such an effect. However, different sources of emission are also possible, i.e outer layers of the secondary component, similar to chromospheric emission observed in SB1 $\zeta$And \citep{zetaAnd}. 
We checked several spectra of objects observed by neighboring fibers and found no signs of emission around $\ha$, therefore its source is unlikely to be back/foreground nebulae, which usually have narrow emission lines. We leave the detailed analysis of emission features for future studies.
\par 
The derived properties suggest that the primary subgiant component is much heavier than secondary red giant. This indicates that there was mass transfer in this system. The red giant expanded and reached its Roche lobe and lost a significant fraction of its mass to the subgiant, which is now a cool semidetached binary system. It is similar to the well known triple system Algol, where the inner pair components have mass ratio $q=4.56\pm0.34$ \citep[][]{algol}, which is very close to that of TYC 2990-127-1. 
The inclination of the orbit is too small for eclipses to be seen, unlike the Algol system. 
If we fix the orbital inclination to the value from the LC fit $i=39.8^{\circ}$ the total mass of the system and orbital radius can be constrained: 
we have  $M_{\rm tot} \sim 1.86~M_\odot,~a\sim 18.89~R_\odot$ (inner system of Algol has $M_{\rm tot} = 4.38\pm0.27~M_\odot,~a=13.65\pm 0.07~R_\odot$ from \citet{sb3algol}). 
This indicates that TYC 2990-127-1 is a close binary with a circular orbit, which is common for binaries with $P\leq 10$ days \citep{2021arXiv210710860K}.
\par
The projected rotational velocities measured from the spectra are relatively high ($\vsini =21,\,31\kms$ with free $\logg_{\rm scd}$ and $\vsini =24,\,25\kms$ assuming tidal synchronization) which can be interpreted as a sign of tidally synchronised rotation in a close binary. The calculated values of $\vsini$ are based on parameters from the LC fit are $\vsini= 21.36,\,22.25\,\kms$. A possible explanation for these miss-match are unaccounted turbulent motions ($\Vmac$), which can cause additional broadening of the spectral lines. The fitting results for a high-resolution APOGEE spectrum, taken at the moments near maximal RV separation support this hypothesis, see Appendix~\ref{sec:payne}. 
\par
The secondary component has a higher metallicity by $\Delta\feh\sim 0.2$ dex, which may be due to mass transfer. In its current stage of evolution we can observe inner metal-rich layers, as original outer metal-poor layers were "sucked away" by the companion. This is supported  by the significant abundance differences derived from APOGEE spectrum in Section~\ref{sec:payne}.

\subsection{Binary evolution simulation}
\begin{figure}
    \centering
    \includegraphics[width=\columnwidth]{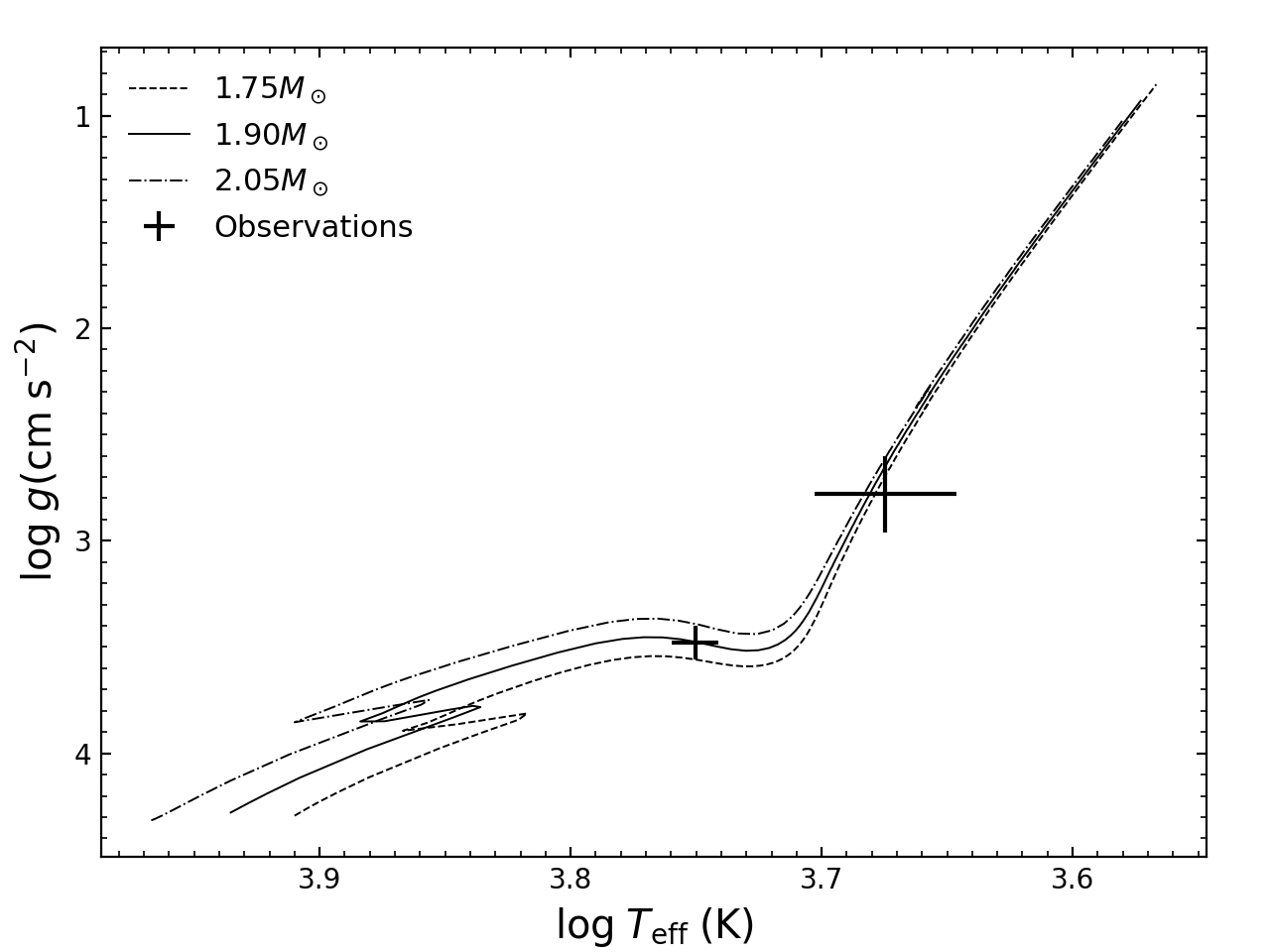}
    \caption{Evolutionary tracks of a single star evolution. The dotted, solid, and dash-dot lines are for stars with mass of $1.75,1.90,2.05M_\odot$, respectively. The observations are shown with error bars, while the primary is the hotter one. }
    \label{fig:8}
\end{figure}

Using the constraints from the observations outlined above, we adopt the binary evolution code MESA (Version 9575, \citealt{paxton2011,paxton2013,paxton2015}) to simulate the evolution of TYC 2990-127-1. The primary is hotter than the secondary and is likely to gain mass from the secondary during the mass transfer phase. In the first step, we limit the primary mass by using the $(\teff,\logg)_{\rm prm}$. In Figure \ref{fig:8}, we perform the single star evolution with mass of $1.75,1.90,2.05M_\odot$, respectively (as shown in dashed, solid and dash-dot lines, respectively). It is clear that the evolutionary tracks cross the primary parameters in the Hertzsprung gap (subgiant stage). Therefore, the primary mass in the current stage is about $1.75\sim2.05 M_\odot$. According to the mass ratio ($4.74\pm 0.09$) in the observation, the secondary mass is estimated to be $0.37\sim0.43M_\odot$. 
\par
\begin{figure}
    \centering
    \includegraphics[width=\columnwidth]{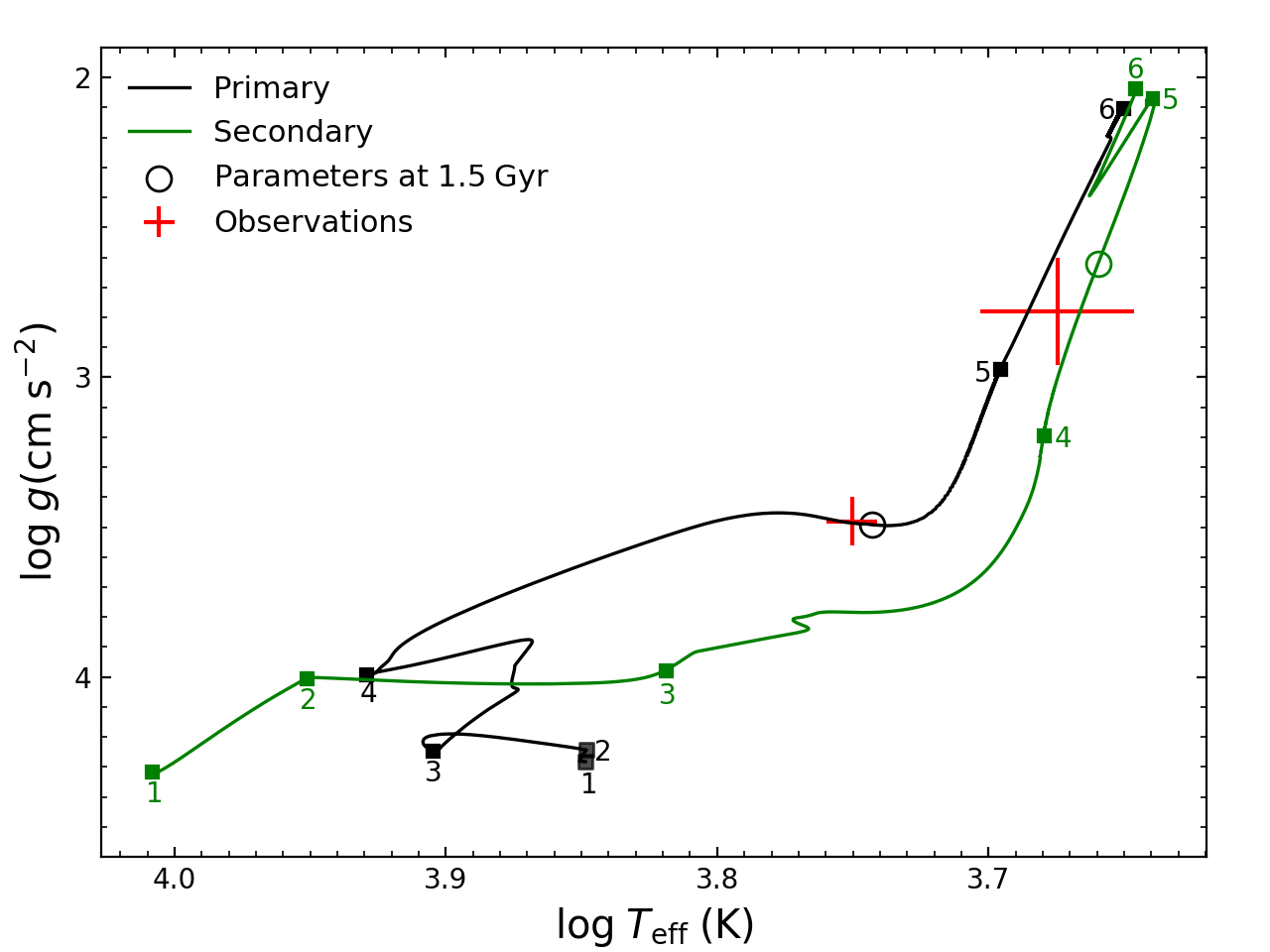}
    \caption{Binary evolution results for the best-fitting model. The initial primary mass, secondary mass, and orbital period are $1.48M_\odot,2.29M_\odot,0.88\;\rm d$, respectively. The accretion efficiency is $0.30$. The evolutionary tracks for the primary and secondary are shown in solid black and green lines, respectively. The open circles are for the binary at the age of $1.5\;\rm Gyr$, when the orbital period, mass ratio, and $(T_{\rm eff},\log\;g)$ of the primary and secondary are within the range of the observation. The evolutionary tracks are calculated from the zero-age main sequence until the primary fills its Roche lobe, points 1 and 6, respectively. See text for more details.}
    \label{fig:9}
\end{figure}

Next, we evolve both stars simultaneously, and generate the progenitor binary parameters. In our best-fitting model, the initial primary (accretor) mass is $1.48M_\odot$ the initial secondary (donor) mass is $2.29M_\odot$, and the initial orbital period is $0.88\;\rm d$. We find a non-conservative mass transfer rate is needed to explain the observation results, and the accretion efficiency is adopted to be $0.30$. The tracks are calculated from zero age main sequence (ZAMS, point 1) until the primary fills its Roche lobe (point 6), as shown in Figure \ref{fig:9}. The secondary evolves first and fills its Roche lobe at $0.41\;\rm Gyr$ (point 2). At the early mass transfer phase, the mass transfer proceeds on a thermal timescale. The primary is rejuvenated by the accreted material, which leads to the increase of effective temperature (from point 2 to 3 as shown in black line). At point 3, the mass ratio reversal happens and the subsequent evolution is driven by nuclear timescale mass transfer. The evolution of the primary is then dominated by its nuclear burning, and it enters into the subgiant stage after point 4 with a mass of $1.98M_\odot$. At this moment, the donor has ascended onto the red giant branch with a mass of $0.61M_\odot$. When the binary evolves to around $1.5\;\rm Gyr$, the binary parameters, e.g. the mass ratio, orbital period and $(\teff,\logg)$ of the primary and secondary, support the observations well. The corresponding results are shown in Table \ref{tab:3}. Here the donor and secondary masses are $0.424$ and $2.041M_\odot$, respectively, and the mass transfer rate is about $2.5\times 10^{-9}M_\odot\;\rm yr^{-1}$. The masses here are larger than that calculated from the LC solution, although this may be due to the overestimated inclination, which is very uncertain. 
The simulation results support that TYC-2990-127-1 is an Algol-like system, and the secondary is transferring mass to the primary.  

The termination of mass transfer is at point 5, where a proto-He WD is left from the donor with a mass of $0.30M_\odot$. Meanwhile, the primary mass is $2.07M_\odot$, and the orbital period is $18.06\;\rm d$. The radius of the primary continues to expand after the end of mass transfer, and the primary will fill its Roche lobe later at point 6. Note that the mass ratio between the primary and secondary is about $7$, the unstable mass transfer is unavoidable in that phase, which results in the common envelope (CE) process. At the onset of the CE phase, the He core mass of primary is $0.30M_\odot$, and the H-rich envelope mass is $1.77M_\odot$. Due to the high uncertainties of the CE process, it is unclear whether the CE can be ejected successfully. The final product after the CE ejection may be a detached double He WD \citep{brown2020b} or a merger, which strongly depends on the CE ejection efficiency \citep{ivanova2013,toonen2013}.

\begin{table}
  \centering
  \caption{Binary parameters at evolutionary age of $1.5\;\rm Gyr$.}
  \begin{tabular}{lcc}
\hline
    & Simulations & Observations \\
\hline
Orbital period (days) & $7.0565$ & $7.0565\pm 0.0004$\\
Mass ratio & $4.81$ & $4.74\pm 0.09$\\
$T_{\rm eff\;prm}$ & $5533$ & $5627\pm 122$\\
$\log(g)_{\rm prm}$ & $3.49$ & $3.48\pm0.08$\\
$T_{\rm eff\;scd}$ & $4566$ & $4727\pm 319$\\
$\log(g)_{\rm scd}$ & $2.62$ & $2.78\pm 0.18$\\
\hline
\end{tabular}
\label{tab:3}
\end{table}

\section{Conclusions}
\label{concl}
We confirm that TYC 2990-127-1 is a spectroscopic binary of SB2 type using medium resolution spectra from the LAMOST MRS survey. We successfully fit a composite spectrum and characterise primary and secondary components. We calculate an accurate orbital solution for this binary and estimate total mass and size of the system. We verify the orbital solution by comparison with a high resolution spectrum taken at a different epoch. This object is a very interesting target for future high-resolution spectroscopic observations and spectral separation techniques \citep{gonzalez2006,sb3algol} and with the help of our orbital solution one can get observations at the moments of optimal RV separation.  
\par
Our results for TYC 2990-127-1 suggest that the methods described in this paper are reliable for estimation of SB2 properties. 
Thus we plan to apply these methods to additional time domain spectra from the LAMOST MRS survey and publish a catalogue of detected spectroscopic binaries in our future paper (Kovalev et al. in prep.).
\par
 The LC from TESS agrees well with the orbital solution. Photometrical and spectroscopic data suggest that TYC 2990-127-1 is a cool Algol-system, where the inclination of the orbit is too small for eclipses to be seen. Significant differences in the masses and surface abundances  suggest, that this system has experienced intensive mass transfer in the past. Mass transfer is probably still active in this system, based on variable emission around H$_\alpha$ and small, irregular variability, seen in the LC.  
\par
The binary evolution simulations suggest that the binary may experience non-conservative mass transfer process at a rate of about $2.5\times 10^{-9}M_\odot\;\rm yr^{-1}$. The results are consistent with previous works that Algol-type systems may be undergoing rapid mass loss (e.g. \citealt{qian2000a,qian2000b,erdem2014}). We find a comparatively lower value of accretion efficiency ($0.3$ adopted in this work) is better to support the observations. We expect that the further observations of the period derivative will put a constraint on the accretion efficiency. The binary will enter the CE phase with the expansion of the secondary, and the remnant product after the CE ejection may be a detached He WD or a merger, depending on the adopted value of the CE ejection efficiency.

\section*{Acknowledgements}
 We are grateful to the anonymous referee for a constructive report. We thank Hans B{\"a}hr for his careful proof-reading of the manuscript.
MK is grateful to his parents, Yuri Kovalev and Yulia Kovaleva, for their full support in making this research possible. The work is supported by the Natural Science Foundation of China (Nos. 11733008, 12090040, 12090043, 11521303, 12125303, 11973053, 11833002,12103086) and Sichuan Science and Technology Program (Grant No. 2020YFSY0034). 
Guoshoujing Telescope (the Large Sky Area Multi-Object Fiber Spectroscopic Telescope LAMOST) is a National Major Scientific Project built by the Chinese Academy of Sciences. Funding for the project has been provided by the National Development and Reform Commission. LAMOST is operated and managed by the National Astronomical Observatories, Chinese Academy of Sciences. The authors gratefully acknowledge the “PHOENIX Supercomputing Platform” jointly operated by the Binary Population Synthesis Group and the Stellar Astrophysics Group at Yunnan Observatories, Chinese Academy of Sciences. 
This research has made use of NASA’s Astrophysics Data System, the SIMBAD data base, and the VizieR catalogue access tool, operated at CDS, Strasbourg, France. It also made use of TOPCAT, an interactive graphical viewer and editor for tabular data \citep[][]{topcat}.
\par

We acknowledge the use of TESS High Level Science Products (HLSP) produced by the Quick-Look Pipeline (QLP) at the TESS Science Office at MIT, which are publicly available from the Mikulski Archive for Space Telescopes (MAST). Funding for the TESS mission is provided by NASA's Science Mission directorate.
\par
This work has made use of data from the European Space Agency (ESA) mission Gaia (\url{https://www.cosmos.esa.int/gaia}), processed by the Gaia Data Processing and Analysis Consortium (DPAC, \url{https://www.cosmos.esa.int/web/gaia/dpac/consortium}). Funding for the DPAC has been provided by national institutions, in particular the institutions participating in the Gaia Multilateral Agreement.
\par
Funding for the Sloan Digital Sky 
Survey IV has been provided by the 
Alfred P. Sloan Foundation, the U.S. 
Department of Energy Office of 
Science, and the Participating 
Institutions. 
SDSS-IV acknowledges support and 
resources from the Center for High 
Performance Computing  at the 
University of Utah. The SDSS 
website is www.sdss.org.
SDSS-IV is managed by the 
Astrophysical Research Consortium 
for the Participating Institutions 
of the SDSS Collaboration including 
the Brazilian Participation Group, 
the Carnegie Institution for Science, 
Carnegie Mellon University, Center for 
Astrophysics | Harvard \& 
Smithsonian, the Chilean Participation 
Group, the French Participation Group, 
Instituto de Astrof\'isica de 
Canarias, The Johns Hopkins 
University, Kavli Institute for the 
Physics and Mathematics of the 
Universe (IPMU) / University of 
Tokyo, the Korean Participation Group, 
Lawrence Berkeley National Laboratory, 
Leibniz Institut f\"ur Astrophysik 
Potsdam (AIP),  Max-Planck-Institut 
f\"ur Astronomie (MPIA Heidelberg), 
Max-Planck-Institut f\"ur 
Astrophysik (MPA Garching), 
Max-Planck-Institut f\"ur 
Extraterrestrische Physik (MPE), 
National Astronomical Observatories of 
China, New Mexico State University, 
New York University, University of 
Notre Dame, Observat\'ario 
Nacional / MCTI, The Ohio State 
University, Pennsylvania State 
University, Shanghai 
Astronomical Observatory, United 
Kingdom Participation Group, 
Universidad Nacional Aut\'onoma 
de M\'exico, University of Arizona, 
University of Colorado Boulder, 
University of Oxford, University of 
Portsmouth, University of Utah, 
University of Virginia, University 
of Washington, University of 
Wisconsin, Vanderbilt University, 
and Yale University.

\section*{Data Availability}
The data underlying this article will be shared on reasonable request to the corresponding author.




\bibliographystyle{mnras}




\appendix

\section{LC modelling using \texttt{W-D} code}
\label{app:lc}
\par
The light curve presents a round, sine-like variations with a low amplitude of about 0.06 mag. The general feature suggests that TYC~2990-127-1 is an ellipsoidal variable rather than an eclipsing binary. We try to analyse the LC using the Wilson–Devinney code \citep[\texttt{W-D}][]{wd71,wilson79} of 2013 version.
Since the TESS bandpass has not been included in the W-D code, we used the Kepler bandpass as a substitute.
According to the behaviour of the radial velocity curves, we set the massive, primary component as Star1 and fixed the mass ratio $Q=1/q=M_{2}/M_{1}=0.21$.  The surface
temperatures of the components were set as $T_{1}=5627$ K and $T_{2}=4727$ K. The bolometric albedos of the two stars were taken to be $A_{1}=A_{2}=0.5$. The gravity darkening exponents were set to $\alpha_{1,2}=0.32$. The initial bolometric limb-darkening coefficients in logarithmic form ($X_{1,2}$) were taken from \citet{hamme1993}, and the monochromatic ones ($x_{1,2}$) were adopted from \citet{claret2017}. The circular orbit ($e=0$) and synchronous rotation for both components' ($F_{1}=F_{2}= 1.0$) assumptions were adopted.  
\par
\begin{figure}
	\includegraphics[width=\columnwidth]{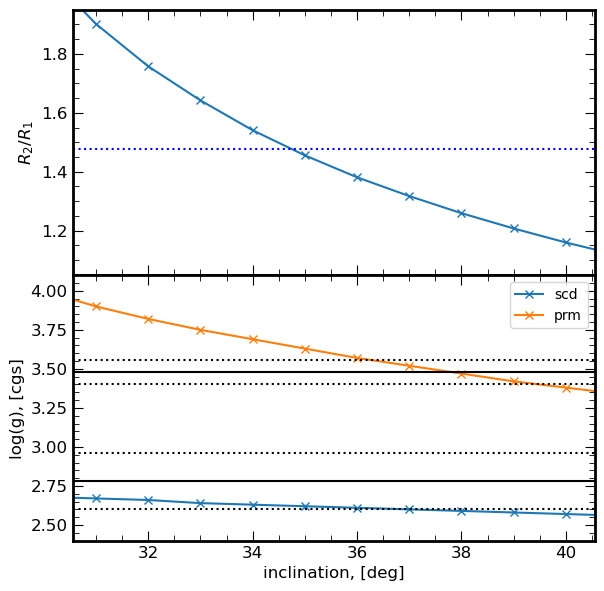}

    \caption{Degeneracy of the model parameters: } ratio of the mean radii (top panel) and surface gravities (bottom panel) versus inclination angle. Values which are shown on y-axis are computed using the best LC fit for a given inclination. Black horizontal lines represent spectroscopic $\logg$, with typical uncertainties, on the lower panel. Blue dotted line represents the ratio of the projected rotational velocities of the top panel.
    \label{fig:inc}
\end{figure}
\begin{figure}

	\includegraphics[width=\columnwidth]{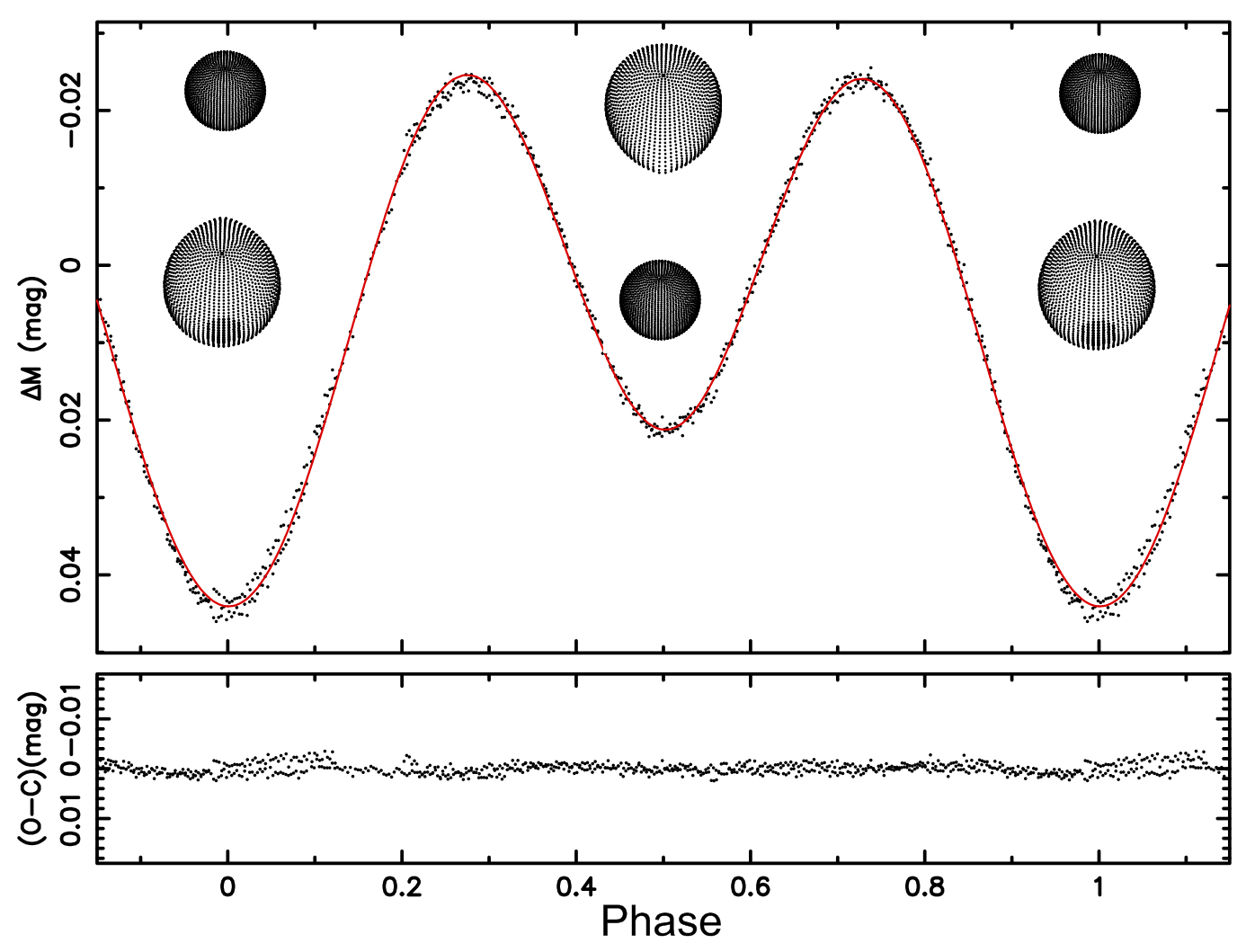}

    \caption{The best LC fit computed for inclination $i=34.5^{\circ}$. The configurations of the system during LC minima are also shown.}
    \label{fig:lc}
\end{figure}

Since the ellipsoidal variations are not sensitive to inclination, it is needed to take the results from spectral analysis as constraints to obtain an approximate solution. We therefore made grids of test solutions with given inclinations from 25 to 40 degrees at the outset. At each assumed inclination, the differential-correction program of the \texttt{W-D} code was started from mode 2 (detached configuration) and rapidly converged into mode 5, confirming the semi-detached nature of binary system, as predicted. To improve the LC fitting we added a cool spot on the secondary and adjusted the spot parameters (including latitude, longitude, radius and temperature fraction $T_{\rm spot}/T_{2}$) along with other free parameters (the potentials and dimensionless luminosities of the two stars). In this way, the best-fit solution under each assumed inclination was reached. The solution with $i=34^{\circ}$ has the minimal sum of the residuals. Based on test solutions, surface gravities $\logg_1,~\logg_2$ and the ratio $R_{2}/R_{1}$ were computed and plotted in Figure~\ref{fig:inc} as functions of the inclination. It suggests that the LC solution around $i=36^{\circ}$ matches the spectral solution for $\logg$ well (taking to account typical errors), although $i=34^{\circ}$ matches the ratio $R_{2}/R_{1}$, which should be equal to the ratio of $\vsini$ assuming spin/orbit synchronisation.  It suggests that the LC solution around $i=34.5^{\circ}$ would match the spectral solution well. 
We therefore set $i=34.5^{\circ}$ and compute the LC solution again. The final results for $i=34.5^{\circ}$ are given in Table~\ref{tab:lc} \footnote{We also tested $\teff$ from the spectral solution with tidal synchronisation, but found almost identical results.}. The light curve synthesis together with system's geometry are illustrated by Figure~\ref{fig:lc}.

We also checked the LC solution using \texttt{PHOEBE} model and find that it is consistent with \texttt{W-D}.

\begin{table}
\centering
\caption{Light-curve solution of TYC 2990-127-1}
\begin{tabular}{lcc}

\hline
\hline
Parameter & Star1 & Star2\\
\hline
$i$,  deg     &\multicolumn{2}{c}{34.5*}\\
$Q$            &\multicolumn{2}{c}{0.21*}\\
$A$            & 0.5*  & 0.5*\\
$\alpha$            & 0.32* & 0.32*\\
$T$, K            & 5627* & 4727*\\
$\Omega$  & 6.114 & 2.258\\
$(L/L_{\rm total})_{\rm TESS}$ & 0.508 & 0.492\\
\hline
Spot parameters:\\
Latitude, deg  & {}  & 81.8\\
Longitude, deg & {}  & 181.7\\
Radius, deg    & {}  & 15.5\\
$T_{\rm spot}/T_{2}$ & {} & 0.837\\
\hline
Physical parameters:\\
$M,\,M_{\odot}$ & 2.16  & 0.45\\
$R,\, R_{\odot}$ & 3.62 & 5.42\\
$\logg$, (cgs)   & 3.66 & 2.63\\
\hline
\hline
*: assigned.
\end{tabular}
\label{tab:lc}
\end{table} 

\par
\begin{figure}
    \centering
    \includegraphics[width=\columnwidth]{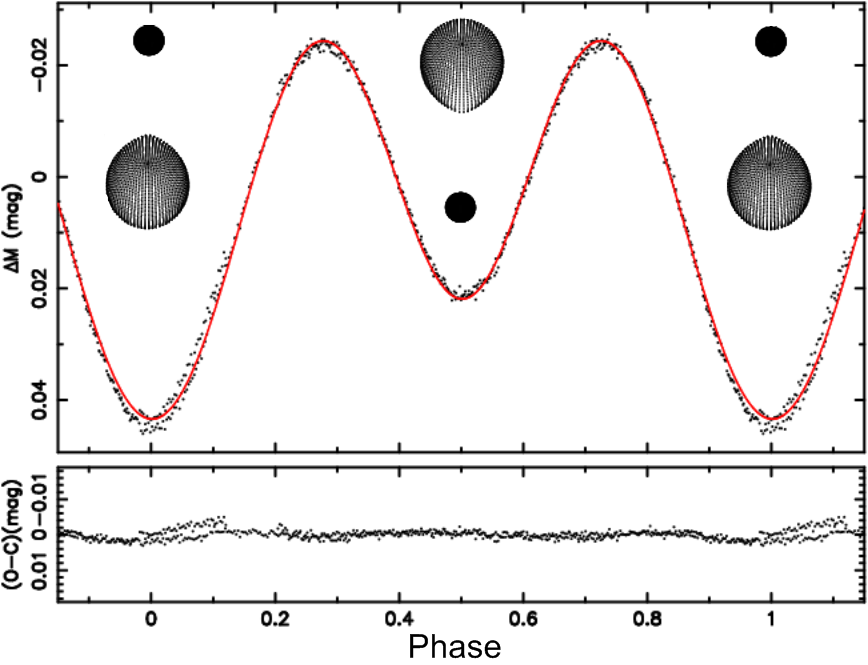}
    \caption{Same as Figure~\protect\ref{fig:lc} but for solution without spot.}
    \label{fig:smallprm}
\end{figure}

 Additionally we present the best LC fit where we did not constrain binary parameters by the spectroscopic solution. In this case, the primary effective temperature $T_1$ and the secondary gravity darkening exponent $\alpha_2$ were fitted. In the best solution, the primary becomes $\sim$ 3 times smaller than the secondary and contributes less light in the TESS band, see Figure~\ref{fig:smallprm}. The parameters from the solution are collected in Table~\ref{tab:lc-spot}.
\begin{table}
\centering
\caption{Light-curve solution without spot.}
\begin{tabular}{lcc}

\hline
\hline
Parameter & Star1 & Star2\\
\hline
$i$,  deg     &\multicolumn{2}{c}{30.72}\\
$Q$            &\multicolumn{2}{c}{0.21*}\\
$A$            & 0.5*  & 0.5*\\
$\alpha$            & 0.32* & 0.04\\
$T$, K            & 5319 & 4727*\\
$\Omega$  & 11.652$\pm$0.091 & 2.257*\\
$(L/L_{\rm total})_{\rm TESS}$ & 0.17 & 0.83\\
\hline
Physical parameters:\\
$M,\,M_{\odot}$ & 3.03  & 0.64\\
$R,\, R_{\odot}$ & 2.09 & 6.06\\
$\logg$, (cgs)   & 4.28 & 2.68\\
\hline
\hline
*: assigned.
\end{tabular}
\label{tab:lc-spot}
\end{table} 

\section{Convective zone evolution in donor star}
\begin{figure}
    \centering
    \includegraphics[width=\columnwidth]{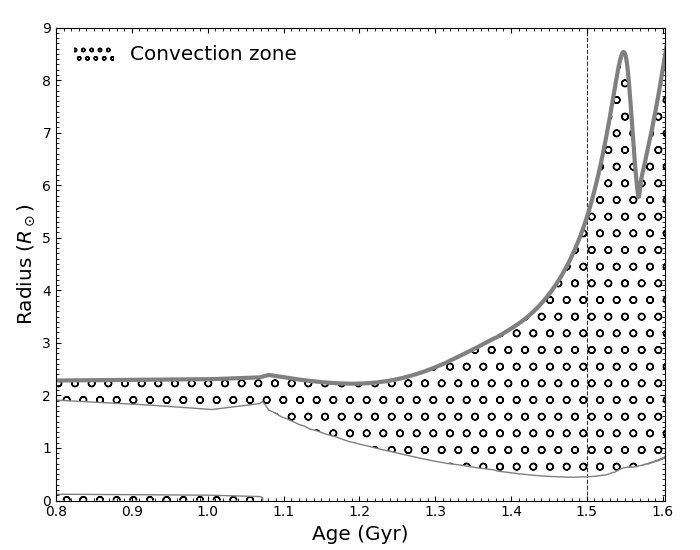}
    \caption{The structure of the convection zone (the shaded region) for the donor star with an initial mass of $2.29M_\odot$, where the evolution of the stellar radius is shown with a grey line and the evolutionary age of $1.5\;\rm Gyr$ is shown in black dashed line. In the later stage of the mass transfer phase, a thick convective envelope has developed for the donor.}
    \label{fig:conv}
\end{figure}

We present the evolution of convective zone (shaded region) for the donor star in Figure \ref{fig:conv}, where the evolution of stellar radius is shown in grey line and the evolutionary age of $1.5\;\rm Gyr$ is shown in black dashed line. The donor star initially has a convective core and a radiative envelope. With the expansion of the envelope, the donor gradually develops a convective envelope and the core becomes radiative during the mass transfer phase. The magnetic field may be generated by the motion of convective envelope according to the dynamo theory \citep{dynamo2014}. 

\section{Fitting the APOGEE spectrum with \textit{the Payne}}
\label{sec:payne}
In this section we attempt to derive chemical abundances from the high-resolution APOGEE spectrum using a binary spectral model based on our orbital solution and single star model from \citet{ting2019}. \textit{The Payne} spectral model\footnote{\url{https://github.com/tingyuansen/The_Payne}} is based on neural networks, which is able to retrieve stellar parameters, the carbon isotope ratio and up to 19 abundances of chemical elements from the infrared spectra of single stars \citep[see][for details]{ting2019}. Unfortunately, rotational broadening is not included in the list of parameters, but there are $\Vmac$ and $\Vmic$. So we cannot simply substitute this model into Equation~\ref{eq:bolzmann}, we have to apply the rotational broadening for each stellar component. We use \texttt{eniric} \citep{jason_neal_2019_2658917} with the standard limb darkening value $\epsilon=0.6$ and $\vsini$ in the range $1:50\,\kms$. We fix the mass ratio $q=q_{\rm dyn}$ and the radial velocities according to the orbital solution. To minimise the number of free parameters we assume tidal synchronisation (Equation~\ref{eq:tidal}). Thus we have a binary spectral model with 51 free parameters.

\par
The spectrum contains many processing artefacts (e.g. telluric lines) and we manually select them and increase the spectral error value for them.  \textit{The Payne} also provides a mask for areas of the spectrum where the solution is incorrect (based on a comparison with the spectra of the Sun and Arcturus). This mask is calculated for the rest-frame spectrum  so we recalculate it according to the velocities of both components and get a new mask. Therefore the new mask has about twice as many wavelength points ($25\%$ vs $12\%$). The spectrum error is set to infinity for masked regions. The initial approximation for the optimiser is taken from our best spectral solution, so that all chemical element abundances are equal to the metallicity $\feh$. The macroturbulence has been constrained ($\Vmac\loa6\kms$) because the main broadening factor in our binary system is rotation. We apply the same APOGEE spectrum normalisation as one provided with \textit{The Payne} model.
\par
\begin{figure}
	\includegraphics[width=\columnwidth]{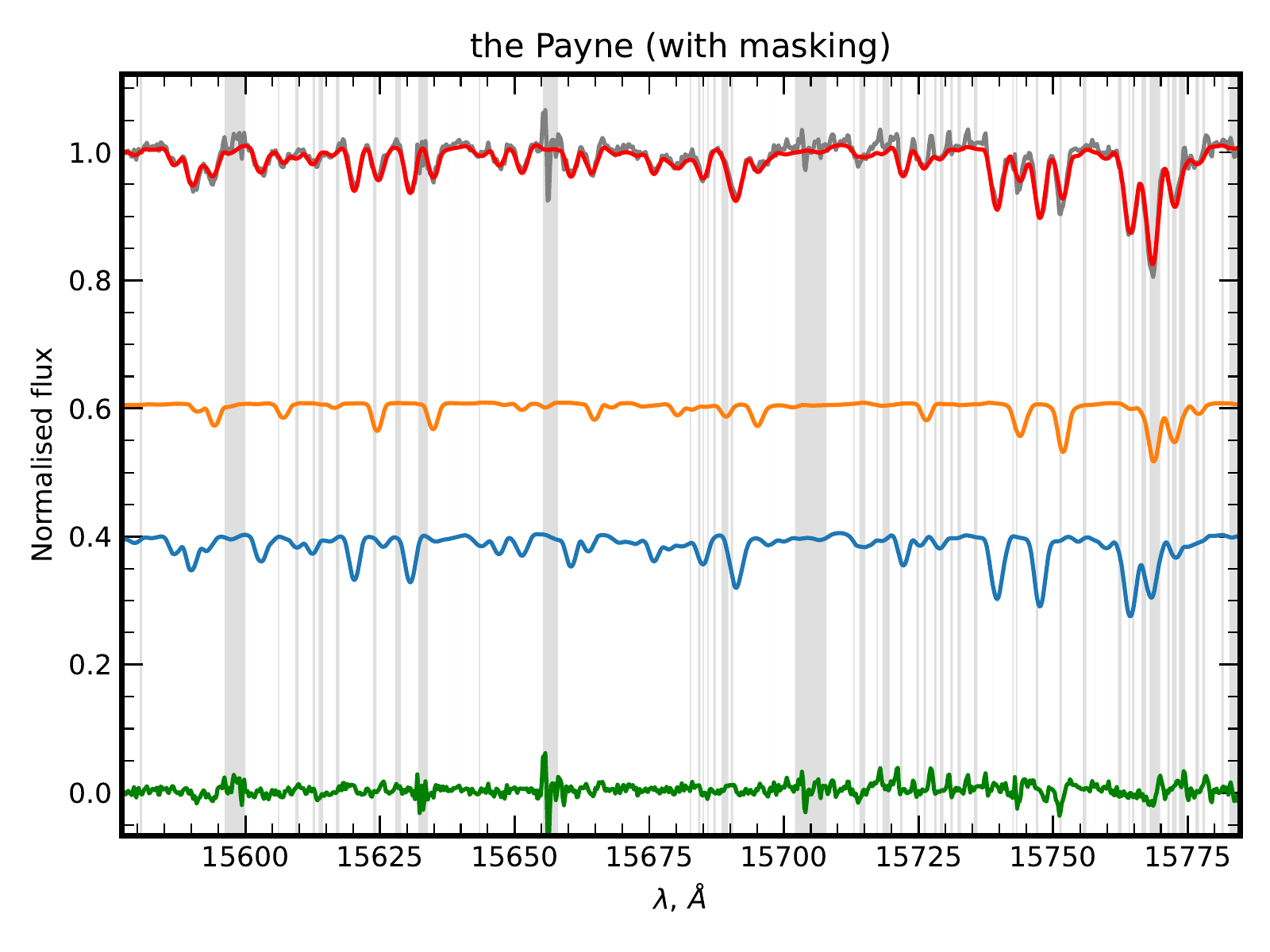}
	\includegraphics[width=\columnwidth]{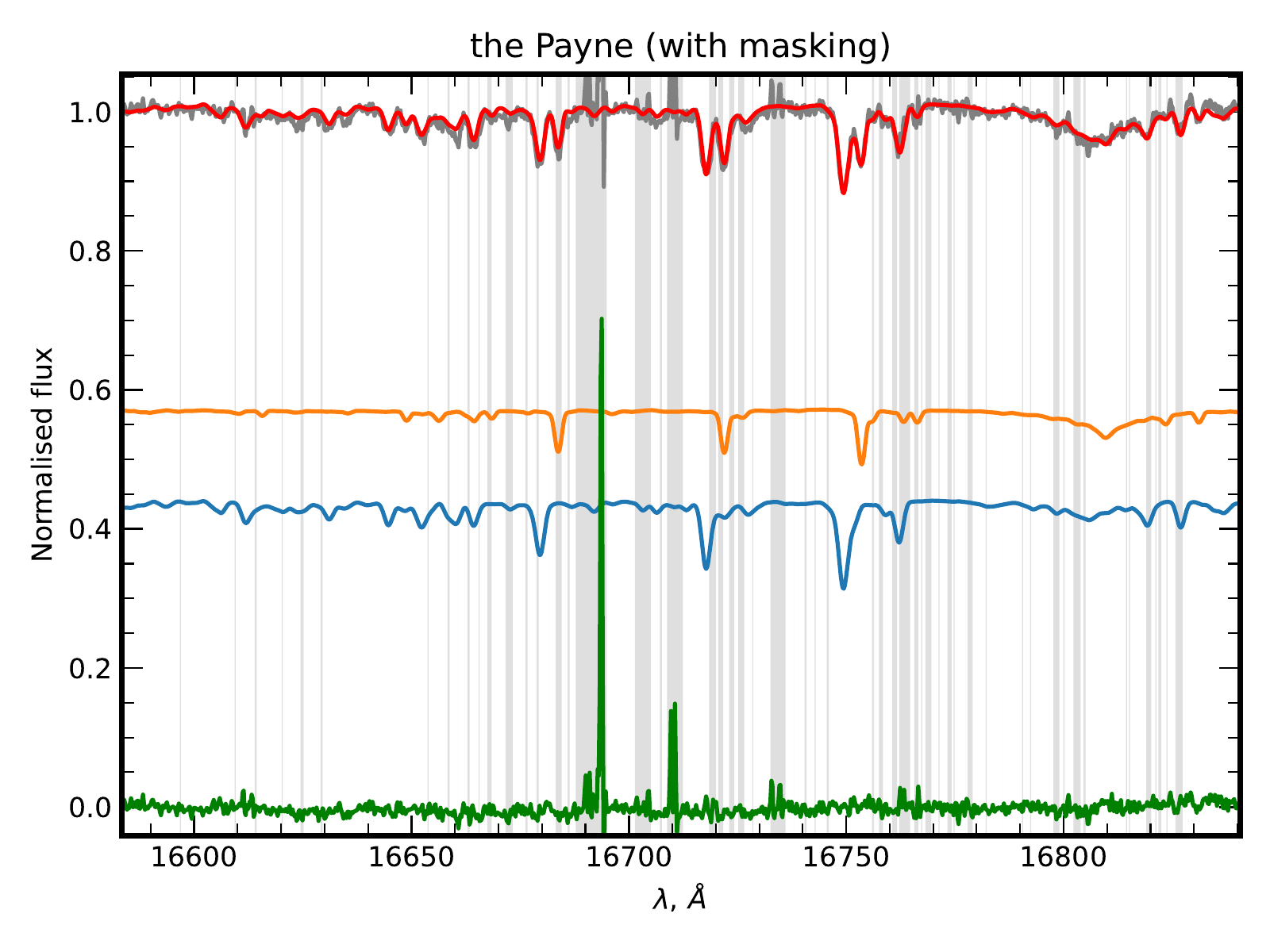}
    \caption{Parts of the normalised APOGEE~DR16 spectrum together with best fit binary model based on \textit{The Payne}. All colours are the same as in Figure~\ref{fig:spfit}. Gray shaded regions were masked during fit.}
    \label{fig:payne}
\end{figure}
\begin{figure}
	\includegraphics[width=\columnwidth]{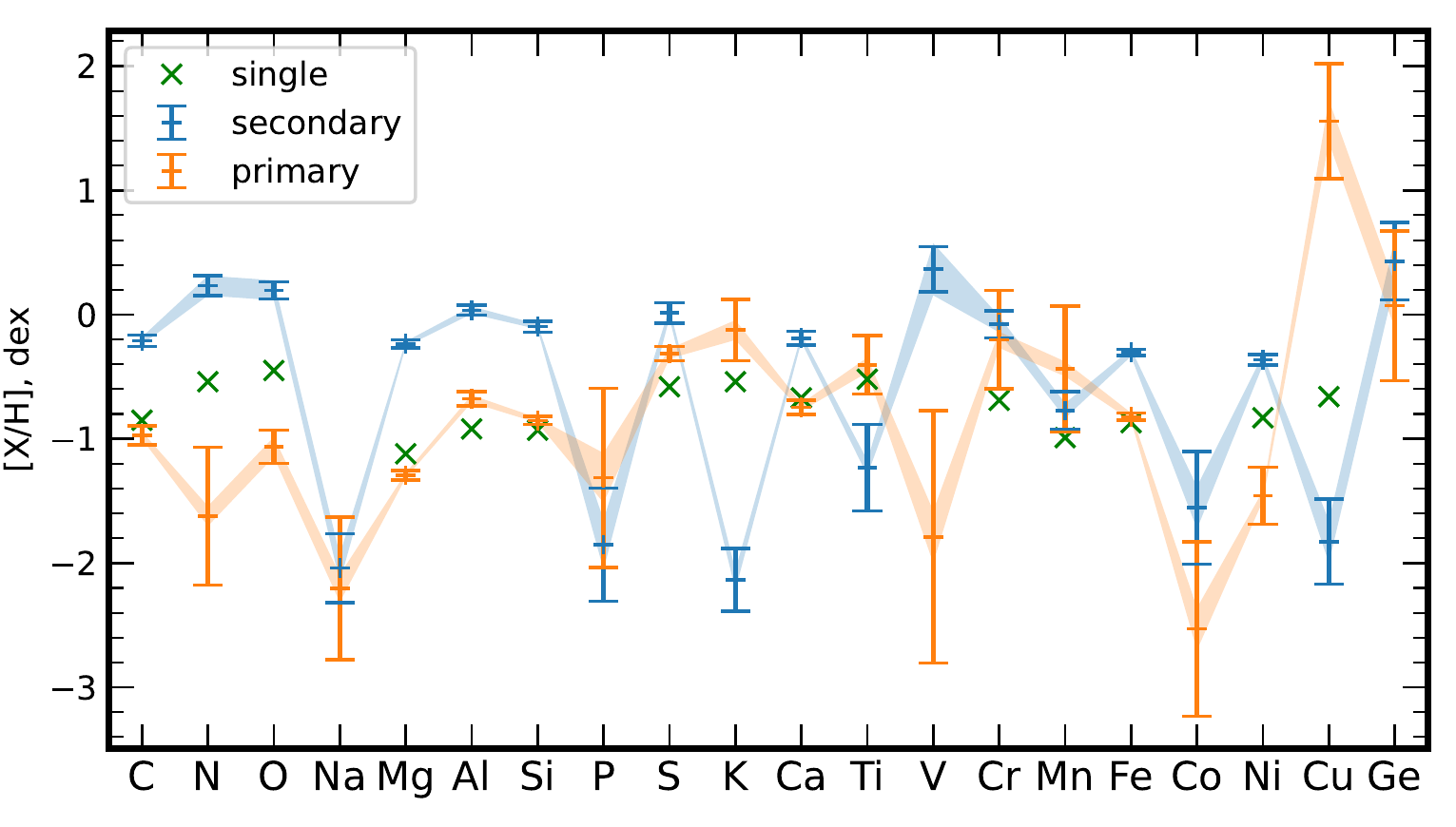}
    \caption{Chemical abundances for primary and secondary components. Shaded regions are internal accuracy of \textit{The~Payne}, green crosses are single star fit results, taken from \protect\citet{ting2019}. Generally, the secondary component has higher abundances ($\Delta=0.33:1.26$ dex) which agrees with intensive mass transfer in the past of the binary system. Original outer metal-poor layers of the donor star were "sucked away" by the companion, while the thick convective zone (see Figure~\protect\ref{fig:conv}) allows for the mixture of the material between the deep metal-rich layers and surface layers}.
    \label{fig:abds}
\end{figure}

In Figure~\ref{fig:payne} we present the best fit of the APOGEE~DR16 spectrum. The agreement is much better than in Figure~\ref{fig:apogee} as \textit{The Payne} allows for a simultaneous fit of many chemical abundances. The derived parameters are collected in Table~\ref{tab:payne} together with single star fit parameters from \citet{ting2019} for comparison. The $\teff,\vsini$ values are very close and $\logg$ values are higher in comparison with values from Table~\ref{tab:final}.  $\Vmac$ are large and close to the maximal allowed value, which indicates that turbulent motions may play a significant role in the line broadening. In Figure~\ref{fig:abds} we plot all derived abundances for the primary and secondary components and values from the single star fit from \citet{ting2019}.  Generally, the secondary component has higher abundances ($\Delta=0.33:1.26$ dex) which agrees with intensive mass transfer in the past of the binary system. Original outer metal-poor layers of the donor star were "sucked away" by the companion, while the thick convective zone (see Figure~\ref{fig:conv}) allows for the mixture of the material between the deep metal-rich layers and surface layers. Abundances of Na, P, K, Ti, V, Co, Cu and Ge are not reliable for both components. In the primary star N, Cr, Mn and Ni abundances are very uncertain. 
Possibly our new mask is not allowed to derive abundances for some of these elements. This can happen if the mask for one stellar component covers the region of the spectrum used by \textit{The Payne} to recover certain element. However we are not able to validate this hypothesis, as it is beyond the scope of this paper. Results of this experiment qualitatively support previous mass transfer in this binary system, however they lack credibility to make quantitative analysis.

\begin{table}
    \centering
    \caption{Derived properties from APOGEE spectrum. $\Delta$ is the difference between the secondary and primary. Single star values are from \protect\cite{ting2019}}
    \begin{tabular}{ccccc}
\hline
\hline
    & secondary & primary & $\Delta$ & single\\
\hline
$\teff$ K & 4718$\pm$40 & 5513$\pm$52& -795$\pm$66 & 4893\\
$\logg$ cgs & 3.53 & 4.17$\pm$0.05 & -0.64$\pm$0.05 & 3.02\\
$\Vmic\,,\kms$ & 2.81$\pm$0.23 & 0.55$\pm$0.66 & 2.26$\pm$0.70 & 0.10\\
\hbox{[C/H]}, dex & -0.21$\pm$0.05 & -0.97$\pm$0.08 & 0.76$\pm$0.09 & -0.85\\
\hbox{[N/H]}, dex & 0.23$\pm$0.08 & -1.62$\pm$0.56 & 1.85$\pm$0.56 & -0.54\\
\hbox{[O/H]}, dex & 0.20$\pm$0.07 & -1.07$\pm$0.13 & 1.26$\pm$0.15 & -0.45\\
\hbox{[Na/H]}, dex & -2.04$\pm$0.28 & -2.20$\pm$0.57 & 0.16$\pm$0.64 & --\\
\hbox{[Mg/H]}, dex & -0.24$\pm$0.03 & -1.29$\pm$0.04 & 1.06$\pm$0.05 & -1.12\\
\hbox{[Al/H]}, dex & 0.04$\pm$0.04 & -0.68$\pm$0.06 & 0.71$\pm$0.07 & -0.92\\
\hbox{[Si/H]}, dex & -0.10$\pm$0.04 & -0.85$\pm$0.03 & 0.75$\pm$0.05 & -0.93\\
\hbox{[P/H]}, dex & -1.85$\pm$0.46 & -1.31$\pm$0.72 & -0.54$\pm$0.85 & --\\
\hbox{[S/H]}, dex & 0.01$\pm$0.08 & -0.31$\pm$0.06 & 0.33$\pm$0.10 & -0.58\\
\hbox{[K/H]}, dex & -2.14$\pm$0.25 & -0.12$\pm$0.25 & -2.01$\pm$0.35 & -0.54\\
\hbox{[Ca/H]}, dex & -0.19$\pm$0.05 & -0.75$\pm$0.06 & 0.56$\pm$0.08 & -0.67\\
\hbox{[Ti/H]}, dex & -1.23$\pm$0.35 & -0.41$\pm$0.24 & -0.83$\pm$0.42 & -0.52\\
\hbox{[V/H]}, dex & 0.37$\pm$0.18 & -1.79$\pm$1.02 & 2.15$\pm$1.03 & --\\
\hbox{[Cr/H]}, dex & -0.08$\pm$0.11 & -0.20$\pm$0.40 & 0.12$\pm$0.41 & -0.69\\
\hbox{[Mn/H]}, dex & -0.77$\pm$0.15 & -0.44$\pm$0.51 & -0.34$\pm$0.53 & -0.99\\
\hbox{[Fe/H]}, dex & -0.31$\pm$0.03 & -0.82$\pm$0.03 & 0.52$\pm$0.04 & -0.87\\
\hbox{[Co/H]}, dex & -1.55$\pm$0.45 & -2.53$\pm$0.70 & 0.98$\pm$0.83 & --\\
\hbox{[Ni/H]}, dex & -0.36$\pm$0.04 & -1.46$\pm$0.23 & 1.09$\pm$0.23 & -0.83\\
\hbox{[Cu/H]}, dex & -1.83$\pm$0.34 & 1.56$\pm$0.46 & -3.39$\pm$0.58 & -0.66\\
\hbox{[Ge/H]}, dex & 0.43$\pm$0.31 & 0.07$\pm$0.60 & 0.36$\pm$0.68 & --\\
$C_{12}/C_{13}$ \% & 83.18$\pm$25.30 & 37.24$\pm$25.13 & 45.95$\pm$35.66 & 44.0\\
$\Vmac\,,\kms$ & 5.98$\pm$1.32 & 5.76$\pm$3.40 & 0.23$\pm$3.65 & 29.99\\
$\vsini,\kms$ & 26$\pm$1 & 25$\pm$1 & 1$\pm$2 & --\\
\hline
\hline
    \end{tabular}
    \label{tab:payne}
\end{table}


\bsp	
\label{lastpage}
\end{document}